\documentclass[fleqn,usenatbib]{mnras}
\usepackage[T1]{fontenc}
\DeclareRobustCommand{\VAN}[3]{#2}
\let\VANthebibliography\thebibliography
\def\thebibliography{\DeclareRobustCommand{\VAN}[3]{##3}\VANthebibliography}
\usepackage{hyperref}
\usepackage{bm}
\usepackage{amsmath}
\usepackage{algorithm}
\usepackage{amsfonts}
\usepackage{mathrsfs}
\usepackage{color}
\usepackage{graphicx} 
\graphicspath{{fig/}}
\usepackage{multicol}
\usepackage{appendix}
\usepackage{url}
\usepackage{soul}

\usepackage{siunitx}

\def\cha{\textit{Chandra}}
\def\XMM{{XMM-{\it{Newton}}}}
\def\NuSTAR{{\it{NuSTAR}}}

\title[NuSTAR Extragalactic Survey of the NEP Field]{The \textbf{\itshape N\MakeLowercase{u}STAR} Extragalactic Survey of the \textbf{\itshape James Webb Space Telescope} North Ecliptic Pole Time-Domain Field}

\author[X. Zhao et al.]{
X. Zhao$^{1,2}$,
F. Civano$^{1}$,
F.~M. Fornasini$^{3,1}$,
D.~M. Alexander$^{4}$,
N. Cappelluti$^{5}$,
C.~T. Chen$^{6}$,
S.~H. Cohen$^{7}$,
\newauthor
M. Elvis$^{1}$,
P. Gandhi$^{8}$,
N.~A. Grogin$^{9}$,
R.~C. Hickox$^{10}$,
R.~A. Jansen$^{7}$,
A. Koekemoer$^{9}$,
G. Lanzuisi$^{11}$,
\newauthor
W.~P. Maksym$^{1}$,
A. Masini$^{12}$,
D.~J. Rosario$^{4}$,
M.~J. Ward$^{4}$,
C.~N.~A. Willmer$^{13}$,
R.~A. Windhorst$^{7}$
\\
\\
$^{1}$Center for Astrophysics $|$ Harvard \& Smithsonian, 60 Garden Street, Cambridge, MA 02138, USA\\
$^{2}$Department of Physics \& Astronomy, Clemson University, Clemson, SC 29634, USA\\
$^{3}$Department of Physics and Astronomy, Stonehill College, 320 Washington Street, Easton, MA 02357, USA\\
$^{4}$Centre for Extragalactic Astronomy, Department of Physics, Durham University, South Road, Durham, DH1 3LE, UK\\
$^{5}$Physics Department, University of Miami, Coral Gables, FL 33124, USA\\
$^{6}$Astrophysics Office, NASA Marshall Space Flight Center, ZP12, Huntsville, AL 35812, USA\\
$^{7}$School of Earth and Space Exploration, Arizona State University, P.O. Box 871404, Tempe, AZ 85287-1404, USA\\
$^{8}$Department of Physics and Astronomy, University of Southampton, Highfield, Southampton, SO17 1BJ, UK\\
$^{9}$Space Telescope Science Institute, 3700 San Martin Drive, Baltimore, MD 21218, USA\\
$^{10}$Department of Physics and Astronomy, Dartmouth College, 6127 Wilder Laboratory, Hanover, NH 03755, USA\\
$^{11}$INAF, Osservatorio di Astrofisica e Scienza dello Spazio di Bologna, via P. Gobetti 93/3, 40129 Bologna, Italy\\
$^{12}$SISSA, Via Bonomea 265, 34151 Trieste, Italy\\
$^{13}$Steward Observatory, University of Arizona, 933 N. Cherry Avenue Tucson, AZ 85721, USA\\
}
\date{Accepted XXX. Received YYY; in original form ZZZ}

\pubyear{2021}

\begin{document}
\label{firstpage}
\pagerange{\pageref{firstpage}--\pageref{lastpage}}
\maketitle

\begin{abstract}
We present the \NuSTAR\ extragalactic survey of the {\it James Webb Space Telescope} ($JWST$) North Ecliptic Pole (NEP) Time-Domain Field. The survey covers a $\sim$0.16\,deg$^2$ area with a total exposure of 681\,ks acquired in a total of nine observations from three epochs. The survey sensitivities at 20\% of the area are 2.39, 1.14, 2.76, 1.52, and 5.20 $\times$ 10$^{-14}$\,erg\,cm$^{-2}$\,s$^{-1}$ in the 3--24, 3--8, 8--24, 8--16, and 16--24\,keV bands, respectively. The NEP survey is one of the most sensitive extragalactic surveys with \NuSTAR\ so far. A total of 33 sources were detected above 95\% reliability in at least one of the five bands. We present the number counts, log$N$-log$S$, measured in the hard X-ray 8--24 and 8--16\,keV bands, uniquely accessible by \NuSTAR\ down to such faint fluxes. We performed source detection on the \XMM\ and \cha\ observations of the same field to search for soft X-ray counterparts of each \NuSTAR\ detection. The soft band positions were used to identify optical and infrared associations. We present the X-ray properties (hardness ratio and luminosity) and optical-to-X-ray properties of the detected sources. The measured fraction of candidate Compton-thick (N$\rm _H\ge10^{24}\,cm^{-2}$) active galactic nuclei, derived from the hardness ratio, is between 3\% to 27\%. As this survey was designed to have variability as its primary focus, we present preliminary results on multi-epoch flux variability in the 3--24 keV band.
\end{abstract}



\section{Introduction}

Active galactic nuclei (AGN), which represent the rapidly growing phase of supermassive black holes \citep[SMBHs;][]{Soltan82,Merloni08}, radiate across the whole electromagnetic spectrum from radio to gamma-ray. X-ray surveys are the most efficient methods to identify and trace the AGN population as accretion processes dominate the emission in the X-ray band. Moreover, X-rays are insensitive to obscuration up to high column density (N$\rm _H\sim$10$^{23}$\,cm$^{-2}$). Thus, X-ray surveys are able to detect a fairly complete AGN population even with large obscuration and faint luminosities. Indeed, multiple X-rays surveys with \cha\ and \XMM\ have been performed in the last 20 years covering a wide range of area and flux \citep{Hasinger_2007, Elvis_2009, Civano_2016, Luo_2016, Chen18, Aneta19}. These works have successfully identified AGN across various redshifts and luminosities, thus measuring the demographics and evolution of both the obscured (N$\rm _H\ge10^{22}\,cm^{-2}$) and unobscured (N$\rm _H<10^{22}\,cm^{-2}$) AGN population up to $z\approx$ 5 \citep{Ueda14,aird15,Miyaji_2015,marchesi2016,Vito17}. 

The Cosmic X-ray Background (CXB), i.e., the diffuse X-ray emission observed between 0.5 and 300\,keV, is produced by the integrated emission from AGN. Indeed, soft X-ray surveys have resolved $\ge$80\% of the CXB into individual AGN at 2--10\,keV and 80--85\% in the 0.5--2\,keV energy range \citep{Hickox06,Cappelluti_2017}. At 20--40 keV, before the launch of the Nuclear Spectroscopic Telescope Array \citep[\NuSTAR;][]{harrison}, only 1\% was resolved by the coded-mask instruments, such as {\it INTEGRAL} and {\it Swift}-BAT \citep{Vasudevan_2013}. \NuSTAR\, the first focusing hard X-ray telescope in orbit, is 100 times more sensitive than the previous high energy instruments, and thus capable to probe AGN up to high redshift ($z\ge$2). Indeed, \NuSTAR\ has resolved $\sim$35--60\% of the CXB at 8--24\,keV \citep[][Hickox et al. in prep]{harrison15}.

As part of the PI guaranteed time, a series of extragalactic surveys were performed by \NuSTAR\ covering from small areas with deep exposures to wide and shallow surveys: the Extended \cha\ Deep Field--South \citep[ECDFS, 0.33\,deg$^2$;][]{Mullaney15}, the Cosmic Evolution Survey field \citep[COSMOS, 1.7\,deg$^2$;][]{Civano_2015}, and the Serendipitous survey \citep[13\,deg$^2$;][Klindt et al. in prep]{Lansbury_2017}. The strategy was then implemented by a few more surveys: the Ultra Deep Survey (UDS) of the UKIRT Infrared Deep Sky Survey \citep[UKIDSS, 0.6\,deg$^2$;][]{Masini_2018}, which further constrained the fraction of Compton-thick (CT-, N$\rm _{H}\ge$ 10$^{24}$\,cm$^{-2}$) AGN in hard X-rays, the Extended Groth Strip survey (EGS, 0.25\,deg$^2$; Aird et al., in prep), and the \cha\ Deep Field--North survey (CDFN, 0.07\,deg$^2$).

During the first call for very large \NuSTAR\ proposals (Cycle 5), our team (PI: Civano) was awarded 585\,ks to observe with \NuSTAR\ a new extragalactic field of interest, the {\it James Webb Space Telescope} \cite[{\it JWST};][]{Gardner:2006ul} North Ecliptic Pole (NEP) Time-Domain Field \citep{Jansen_2018}.

This field was selected by \citet{Jansen_2018} to be located within {\it JWST}’s northern continuous viewing zone (CVZ). {\it JWST} Interdisciplinary Scientist (IDS) R.~Windhorst allocated $\sim$47 hours of his guaranteed time (program JWST-GTO-1176), to be executed in Cycle~1 after launch and on-orbit validation of {\it JWST}. The {\it JWST} NEP Time-Domain Field has become a new and promising field for both time domain studies and population studies, due to its low Galactic foreground extinction, absence of bright foreground stars, and continuous visibility. Indeed, during the past few years, broad community interest has rapidly grown the {\it JWST} NEP Time-Domain Field into a comprehensive multi-wavelength survey. The multi-wavelength coverage of {\it JWST} NEP Time-Domain Field\footnote{\url{http://lambda.la.asu.edu/jwst/neptdf/}} obtained or approved to date includes: {\it JWST}/NIRCam+NIRISS (PI: Windhorst \& Hammel), {\it HST}/WFC3+ACS (PI: Jansen), {\it LBT}/LBC (PI: Jansen), {\it Subaru}/HSC (PI: Hasinger \& Hu), {\it J-PAS} (PI: Bonoli \& Dupke), {\it GTC}/HiPERCAM (PI: Dhillon), {\it TESS} (PI: Berriman \& Holwerda), {\it MMT}/MMIRS+Binospec (PI: Willmer), {\it JCMT} (PI: Smail \& Im), {\it IRAM}/NIKA 2 (PI: Cohen), {\it VLA} (PI: Windhorst \& Cotton), {\it VLBA} (PI: Brisken), {\it LOFAR} (PI: Van Weeren), \cha\ (PI: Maksym). 

In particular, hard X-ray coverage of infrared (IR) fields, which are known to be unbiased to detecting even the most heavily obscured CT-AGN \citep[see a recent review,][]{Padovani:2017we}, is important for understanding the properties (e.g., the obscuration) of the IR detected AGN \citep[e.g.,][]{Gandhi09}. The approved 8-band {\it JWST}/NIRCam imaging and all the photometric and spectroscopic data already available will further constrain the nature of the galaxy hosts, and yield accurate redshifts of hard X-ray detected sources.

In this paper, we present the 0.16\,deg$^2$ \NuSTAR\ extragalactic survey in the {\it JWST} NEP Time-Domain Field. The paper is organized as follows. In Section~\ref{sec:data}, we describe the \NuSTAR\ observations and data processing. In Section~\ref{sec:simulation}, we describe the simulations and the resulting reliability and completeness of our source detection technique. In Section~\ref{sec:catalog}, we present the X-ray properties of the sources detected in the \NuSTAR\ NEP survey. In Section~\ref{sec:multi}, we present the soft X-ray, optical and IR counterparts of the \NuSTAR\ detected sources and their properties. In Section~\ref{sec:discuss}, we discuss the CT fraction and X-ray variability of the sources detected in the \NuSTAR\ NEP survey. In the Appendix, we present the point source catalog. Uncertainties are quoted at 90\% confidence level throughout the paper unless otherwise stated. Standard cosmological parameters are adopted as follows: $\langle H_0\rangle$ = 70 km s$^{-1}$ Mpc$^{-1}$, $\langle\Omega_M\rangle$ = 0.3 and $\langle\Omega_\Lambda\rangle$ = 0.7. 
\begingroup
\renewcommand*{\arraystretch}{1.5}
\begin{table}
\caption{Details of the Individual \NuSTAR\ Observations}
\centering
\label{Table:obs}
\vspace{.1cm}
  \begin{tabular}{ccccc}
       \hline
       \hline     
 	ObsID&Date&R.A.&Dec.&Exp.\\
	&&(deg)&(deg)&(ks)\\
	\hline
	60511001002&2019-09-30&260.8728&65.8221&73.5\\
	60511002002&2019-10-02&260.7727&65.8220&77.6\\
	60511003002&2019-10-04&260.6360&65.8246&68.7\\
	60511004002&2020-01-03&260.5416&65.9071&89.8\\
	60511005002&2020-01-04&260.7314&65.8736&84.7\\
	60511006002&2020-01-05&260.9080&65.9067&83.0\\
	60511007002&2020-03-01&260.5283&65.7657&65.2\\
	60511008001&2020-03-02&260.7546&65.7483&70.2\\
	60511009001&2020-03-03&260.9508&65.7456&68.3\\
	\hline
	\hline
	\vspace{0.06cm}
\end{tabular}
\end{table}
\endgroup

\section{data processing} \label{sec:data}
The \NuSTAR\ survey of the {\it JWST} NEP Time-Domain Field consists of 9 observations with a total exposure time of $\sim$681\,ks acquired during the \NuSTAR\ GO cycle 5 (PI: Civano). The survey was designed to have variability as its primary focus, therefore the field was observed in three epochs during October 2019, January 2020, and March 2020. Details about each observation are listed in Table~\ref{Table:obs}.

\begin{figure} 
\centering
\includegraphics[width=.48\textwidth]{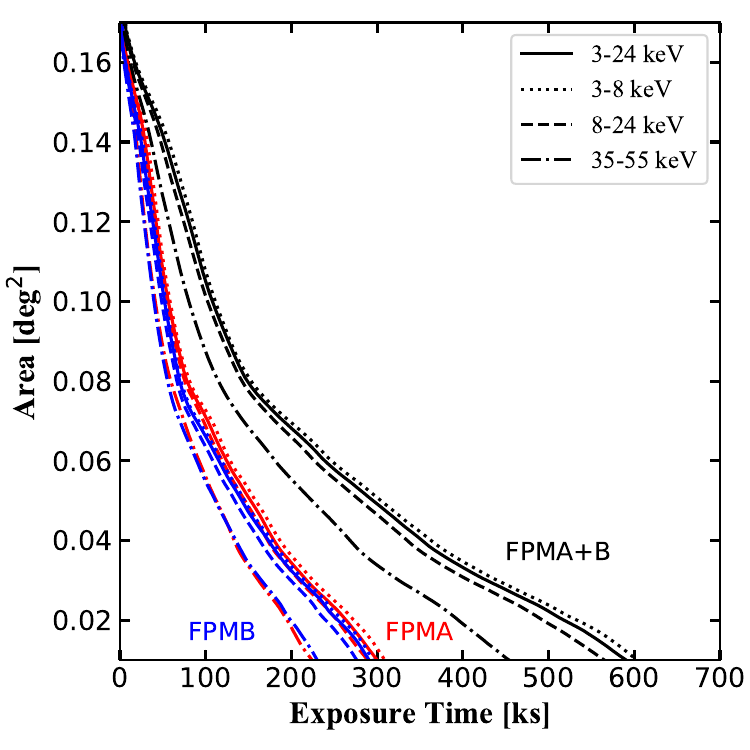}
\caption{Cumulative survey area as a function of the effective vignetting-corrected exposure time for FPMA (red), FPMB (blue), and the summed FPMA+B (black) in four energy bands: 3--24\,keV (solid line), 3--8 keV (dotted line), 8--24 keV (short dashed line), and 35--55\,keV (dash-dotted line).}
\label{fig:expomap}
\end{figure}   

\begin{figure*} 
\begin{minipage}[b]{.49\textwidth}
\centering
\includegraphics[width=\textwidth]{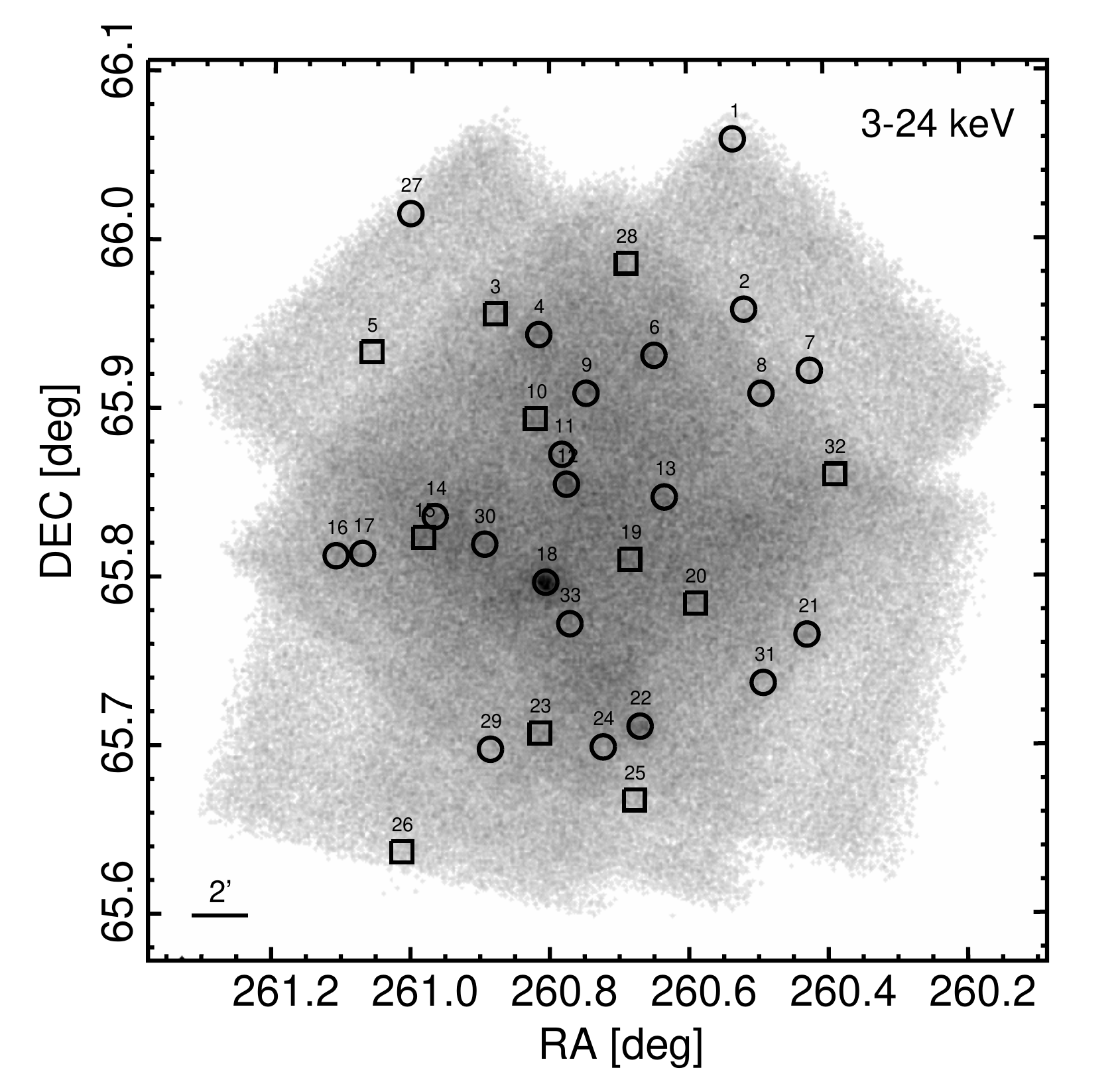}
\end{minipage}
\begin{minipage}[b]{.49\textwidth}
\centering
\includegraphics[width=\textwidth]{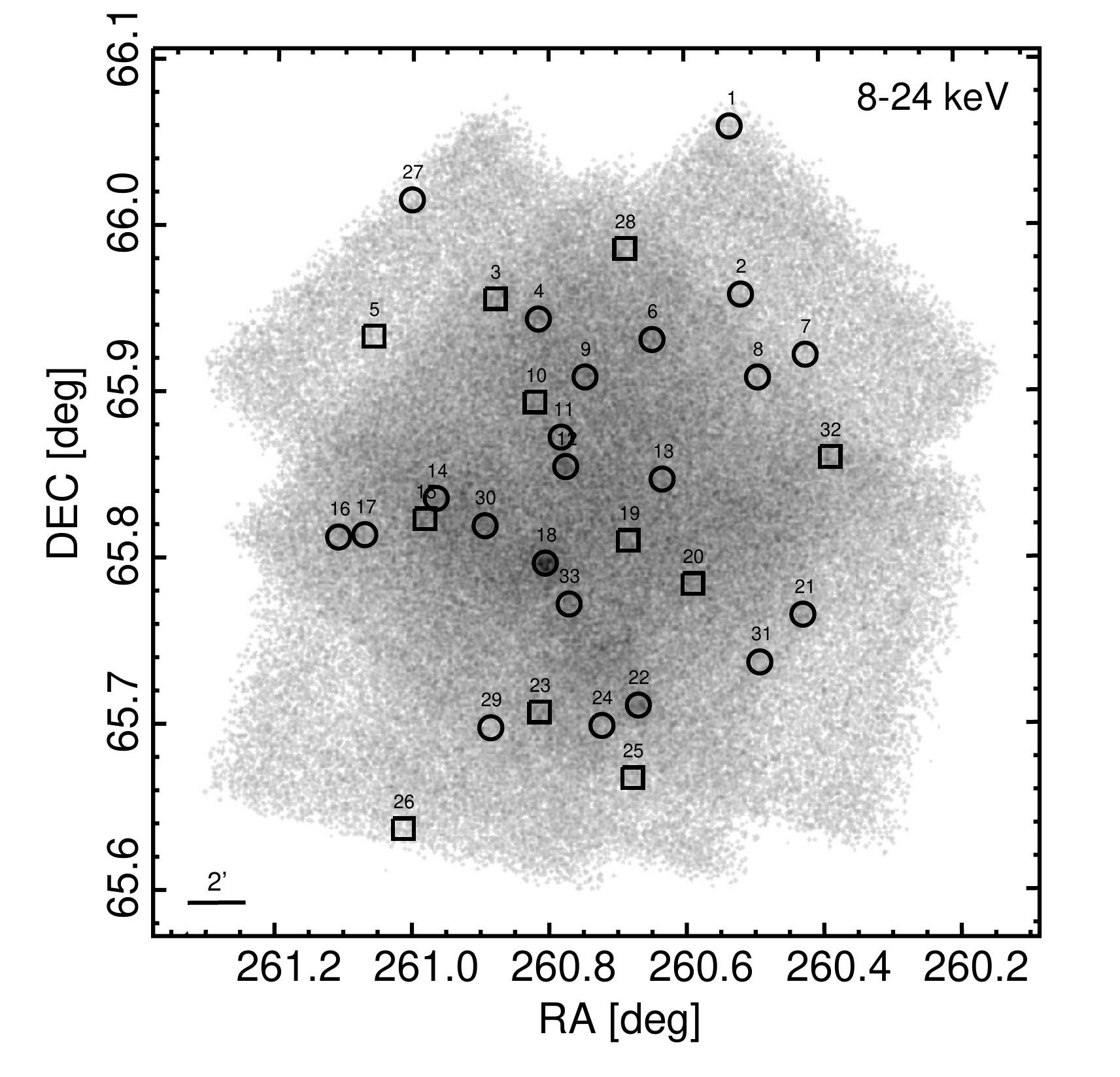}
\end{minipage}
\caption{\NuSTAR\ FPMA+FPMB 3--24\,keV (left) and 8--24\,keV (right) mosaics. The positions of the \NuSTAR\ detected sources with (black circle, 25\arcsec\ radius) and without (black square, 45\arcsec\ width) soft X-ray counterparts are plotted.}
\label{fig:mosaic}
\end{figure*}   

\subsection{Data Reduction}
The \NuSTAR\ data were reduced using HEASoft v.6.27.2 and \NuSTAR\ Data Analysis Software (NuSTARDAS) v.1.9.2 with CALDB v.20200526. The level 1 raw data were calibrated, cleaned, and screened by running the \texttt{nupipeline} script. 

Following \citet[][hereafter, C15]{Civano_2015}, we produced the full-field light curves in the low energy band (3.5--9.5\,keV) with a bin size of 200\,s using \texttt{xselect} to search for high-background time intervals. We screened all nine observations and cleaned the intervals with count rates that were more than a factor of $\sim$2 higher than the average count rate in that observation. The resulting loss of exposure time corresponds to 0.4\%--2.6\% of the individual exposure times of the 9 observations. The total exposure loss of the two \NuSTAR\ focal plane modules are 9.3\,ks for FPMA and 7.0\,ks for FPMB, which is $\sim$1.2\% of the total \NuSTAR\ NEP survey exposure time. We then re-ran \texttt{nupipeline} applying the good time intervals (GTI) after cleaning the high-background time intervals. Due to the strong instrumental emission lines between 25 and 35\,keV, we imposed a high energy cut at 24\,keV in our analysis, which is the standard limit adopted in previous \NuSTAR\ survey studies. Therefore, we filtered the data products of this work into six energy bands (3--24, 3--8, 8--24, 8--16, 16--24, and 35--55\,keV).

\subsection{Exposure Map Production}
The exposure maps were generated in four energy bands 3--24, 3--8, 8--24, and 35--55\,keV using the NuSTARDAS tool \texttt{nuexpomap}, which computes the effective exposure time for each pixel on the detectors. We adopted the 8--24 keV exposure map as the exposure maps of 8--16\,keV and 16--24\,keV energy bands due to the marginal differences between their exposure maps. By using the 8--24\,keV exposure map, the 8--16\,keV exposure is underestimated by at most 3\% and the 16--24\,keV exposure is overestimated by at most 12\% \citep{Masini_2018}. The created exposure maps account for detector bad and hot pixels, detector gaps, attitude variations, mast movements, and telescope vignetting. Since vignetting is heavily energy-dependent, the exposure map of each desired energy band was weighted assuming a power-law spectrum with photon index $\Gamma$ = 1.80. The mean, spectrally weighted energies are 9.88\,keV for 3--24\,keV band, 5.42\, keV for 3--8\,keV band, 13.02\,keV for 8--24 keV band, and 44.35\,keV for 35--55\,keV. The exposure maps were produced using a bin size of 5 pixels (\ang[angle-symbol-over-decimal]{;;12.3}) to reduce the calculation time. The exposure map binning is significantly smaller than the \NuSTAR\ point-spread function (PSF, half power diameter $\approx$ 60\arcsec) and is smaller than the aperture size that we use to perform photometry analysis. The vignetting corrected exposure maps for each observation were then merged into a mosaic for each energy band using the HEASoft tool \texttt{Ximage}. The cumulative area as a function of the effective exposure time for each energy band and for different focal plane modules (FPMA, FPMB, and the summed FPMA+B) is plotted in Figure~\ref{fig:expomap}. A small difference of $<$\,8\% in the effective exposure time is seen between FPMA and FPMB, with FPMA being more sensitive, which is consistent with the expectations.

\subsection{Mosaic Creation}
We merged the nine observations into three mosaics (FPMA, FPMB, and the summed FPMA+B) using the HEASoft tool \texttt{Ximage}. The mosaics are filtered into six energy bands 3--24, 3--8, 8--24, 8--16, 16--24, and 35--55\,keV using the \texttt{Xselect} tool. We did not perform any astrometric offset correction to our data since only one bright object is observed in the \NuSTAR\ NEP field, which prevents us from testing the significance of the astrometric offset as has been done in the previous surveys. C15 found a 1\arcsec--7\arcsec\ astrometric offset in the \NuSTAR\ COSMOS survey using a stacking technique. Therefore, we assume that the astrometric offset only marginally affects the sensitivity of our final mosaic images. The mosaics in the 3--24 and 3--8\,keV bands are presented in Fig.~\ref{fig:mosaic}. To achieve the deepest sensitivity, we performed source detections and simulations using the summed FPMA+B mosaic in each energy band.

\subsection{Background Map Production}
\label{sec:background}
\begin{figure} 
\centering
\includegraphics[width=.5\textwidth]{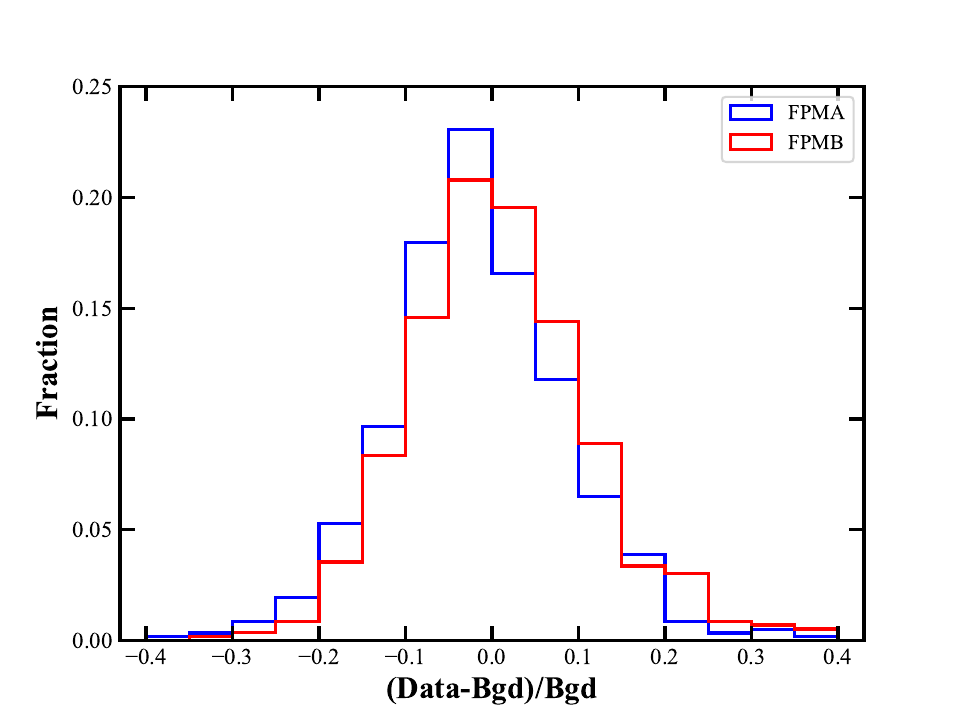}
\vspace{-0.3cm}
\caption{Normalized distributions of relative differences between counts extracted from observed data and background maps in each observation for FPMA (red) and FPMB (blue).}
\label{fig:simulation_quality}
\end{figure}

The \NuSTAR\ background includes four independent components that vary spatially across the detectors and the two telescopes. Here we briefly introduce the different background components \citep[refer to][for more details]{Wik_2014}. Below $\sim$20\,keV, the background is dominated by stray light from the unblocked sky through the aperture stops of \NuSTAR, which is spatially non-uniform across the field of view (FoV). Below $\sim$10\,keV, the X-rays from the CXB which are focused by the optics (fCXB), also contribute to the \NuSTAR\ background. The fCXB component is spatially non-uniform among the four detectors. The solar photons reflected by the back of the aperture stop can also contribute to the background at low-energy ($<$5\,keV). This solar component is time variable depending on whether or not the instrument is in sunlight. Since the NEP survey was conducted during the solar minimum, our observations were not significantly affected by this solar component. Above 20\,keV, the background is dominated by instrumental emission lines produced by interactions between the detectors and radiation in-orbit. We analyzed the background of each \NuSTAR\ observation using {\tt nuskybgd}\footnote{\url{https://github.com/NuSTAR/nuskybgd}} \citep{Wik_2014}, which includes all the aforementioned components. We followed the same procedure as in C15 to produce the background of each observation for both FPMA and FPMB. To ensure the accuracy of the generated background maps, we extract and compare the number of counts of the observations and their corresponding simulated maps. We evenly divided the field of view into 64 circular (45\arcsec\ in radius) subregions. The mean difference between the number of counts are (Data -- Bgd)/Bgd = --0.7\% and 2.3\% for FPMA and FPMB, respectively. The standard deviations of the percent differences are 12.6\% and 14.2\% for FPMA and FPMB, respectively. The normalized distribution of (Data -- Bgd)/Bgd of FPMA and FPMB are plotted in Fig.~\ref{fig:simulation_quality}, which shows a good agreement between the generated background maps and the observed data.

\section{Simulations} \label{sec:simulation}
To test and optimize our source detection method, following C15, we perform an extensive set of simulations (1200 for each energy band\footnote{To build enough statistics, we performed a larger number of simulations than C15 (300 for each energy band) as the area covered by \NuSTAR\ NEP is 14 times smaller than \NuSTAR\ COSMOS.}). Comprehensive simulations are essential to determine the reliability and completeness of the source list and to measure the positional accuracy of the detections.

\subsection{Generating Simulated Data}
To produce the simulated maps, we used the same strategy as in C15. We generated a sample of mock sources with fluxes assigned randomly from the log$N$--log$S$ distribution measured in \citet{Treister09}. The minimum flux for the input sources is set to be 3 $\times$ 10$^{-15}$\,erg\,cm$^{-2}$\,s$^{-1}$ in the 3--24 keV band (about 10 times fainter than the expected flux limit of the survey) to ensure the completeness of the simulation. Fluxes in the 3--8, 8--16, 8--24, and 16--24\,keV bands are extrapolated from the 3--24\,keV flux using an absorbed power-law spectral model with $\Gamma$ = 1.80 and a Galactic absorption of N$\rm _H$ = 3.4 $\times$ 10$^{20}$\,cm$^{-2}$ \citep{nh}. The fluxes of the simulated sources were converted to count rates using count-rate-to-flux conversion factors (CF) of 4.86, 3.39, 7.08, 5.17, and 16.2 $\times$ 10$^{-11}$\,erg\,cm$^{-2}$ counts$^{-1}$ in the 3--24, 3--8, 8--24, 8--16, and 16--24\,keV bands, respectively. The CF were computed using WebPIMMS\footnote{\url{https://heasarc.gsfc.nasa.gov/cgi-bin/Tools/w3pimms/w3pimms.pl}} assuming the above spectral model. 

\begingroup
\renewcommand*{\arraystretch}{1.5}
\begin{table*}
\caption{Mean number of the detected sources using \textit{SExtractor} in the simulated maps (line 1). Mean number of the detected sources matched to the input source catalog within 30\arcsec\ (line 2). The DET\underline{\;\;}ML threshold at 99\% reliability (line 3--5) and at 95\% reliability (line 6--8) for each exposure intervals. Mean number of the detected sources with DET\underline{\;\;}ML above 99\% and 95\% reliability threshold in the simulated maps (line 9,10) and in the real data (line 11,12).}
\centering
\label{Table:number}
  \begin{tabular}{lccccc}
       \hline
       \hline     
 	&3-24\,keV&3-8\,keV&8-24\,keV&8-16\,keV&16-24\,keV\\
	\hline
	Detections in Simulated Maps&393&186&191&184&180\\
	Matched to Input&50&46&37&37&21\\
	\hline 
	DET\underline{\;\;}ML(99\%, 20--200\,ks) threshold&13.53&13.95&15.48&14.43&16.35\\
	DET\underline{\;\;}ML(99\%, 200--500\,ks) threshold&12.00&12.18&13.75&13.10&15.94\\
	DET\underline{\;\;}ML(99\%, $>$500\,ks) threshold&10.99&11.02&13.13&12.30&15.72\\
	\hline 
	DET\underline{\;\;}ML(95\%, 20--200\,ks) threshold&11.47&11.65&12.99&12.35&15.35\\
	DET\underline{\;\;}ML(95\%, 200--500\,ks) threshold&9.89&10.10&11.54&10.88&14.42\\
	DET\underline{\;\;}ML(95\%, $>$500\,ks) threshold&8.89&8.89&10.67&10.22&14.05\\
	\hline 
	Simulated Maps&&&&&\\
	DET\underline{\;\;}ML$>$99\% reliability threshold&14.4&11.4&4.8&5.6&0.5\\
	DET\underline{\;\;}ML$>$95\% reliability threshold&21.1&17.0&7.4&8.5&0.6\\
	\hline 
	Real Data&&&&&\\
	DET\underline{\;\;}ML$>$99\% reliability threshold&16&15&4&10&1\\
	DET\underline{\;\;}ML$>$95\% reliability threshold&26&19&8&11&1\\
	\hline 
	\hline
\end{tabular}
\end{table*}
\endgroup

The simulated maps were created by randomly placing the generated mock sources on the background mosaic produced in Section~\ref{sec:background} but without an fCXB component (since the fCXB is made up of the emission from the undetected discrete sources). Mock source counts were computed using their count rates and the effective exposure at the source location. The counts associated with each mock source were spatially distributed on the background map according to the PSF. The emission from mock sources accounts for only a fraction of the fCXB component. Thus, to match the total number of counts in the observed NuSTAR mosaic, we randomly distributed additional background counts ($<$9\% of the total) on the simulated maps, which were drawn from the background map including only the fCXB component. The fCXB normalization was averaged across all the fields, and re-scaled to account for the ``missing” counts associated with the unresolved fCXB. These additional counts are due to the fact that the unresolved sources fainter than 3 $\times$ 10$^{-15}$ erg\,cm$^{-2}$\,s$^{-1}$ were not included in the simulations. A Poisson realization of the simulated map (mock sources plus background) was generated. With the above procedure, we generated a set of 1200 simulated maps in five energy bands (3--24, 3--8, 8--24, 8--16, and 16--24\,keV) for both FPMA and FPMB, which were then combined to create the FPMA+B simulated maps. The position, exposure time, and flux of each mock source were recorded in an input source catalog for each simulation.

\subsection{Source Detection on Simulated Maps}
\label{sec:detection}
Once we made the simulated maps with mock sources, we performed source detection on the simulated maps, following the technique developed in \citet{Mullaney15}. We performed the source detection on the false-probability maps, which measure the possibility that a signal is due to a background fluctuation rather than from a real source. The false-probability maps were generated with the smoothed simulated maps and the background mosaics using the incomplete Gamma function. The false-probability at each pixel on the map was calculated using the \texttt{igamma} function in IDL, i.e., $P_{\rm false}$ = \texttt{igamma}($C\rm_{data}$, $C\rm_{bgd}$), where $C\rm_{data}$ is the number of counts of the simulated data in that pixel and $C\rm_{bgd}$ is the number of counts of the corresponding background mosaics in that pixel:

\begin{equation}\label{eq:igamma}
P_{\rm false}=\frac{\int_0^{C_{\rm bgd}}{e^{-x}x^{C_{\rm data}-1}{\rm d}x}}{\int_0^\infty{e^{-x}x^{C_{\rm data}-1}{\rm d}x}}
\end{equation}

Before creating the false-probability maps, we smoothed the simulated maps and the background mosaics using circular top-hat functions with 20\arcsec\ radius to ensure the detection of faint sources. We used \textit{SExtractor} \citep{Bertin1996} to detect the sources on the obtained smoothed false-probability maps in each energy band. The false-probability was converted to --\,log($P_{\rm false}$) as the input to \textit{SExtractor}. The detection limit was set to $P_{\rm false}$ $\le$10$^{-2.5}$ ($\sim3\,\sigma$) to make sure that we detected all potentially faint sources. The local minimum $P_{\rm false}$ positions are recorded as the coordinate of each detection. We then define the maximum likelihood (DET\underline{\;\;}ML) for each detection using the inverse of the logarithm of the Poisson probability that a detection is due to a random fluctuation of the background ($P_{\rm random}$), i.e., DET\underline{\;\;}ML = --\,ln\,$P_{\rm random}$. The Poisson probability is calculated using the total and background counts extracted from the simulated map and the corresponding background map at the coordinate of the detection using a circular aperture of 20\arcsec\ radius. A lower DET\underline{\;\;}ML suggests a higher chance that the signal is from the background fluctuation at that position. 

\begin{figure*} 
\centering
\begin{multicols}{2}
\includegraphics[width=.5\textwidth]{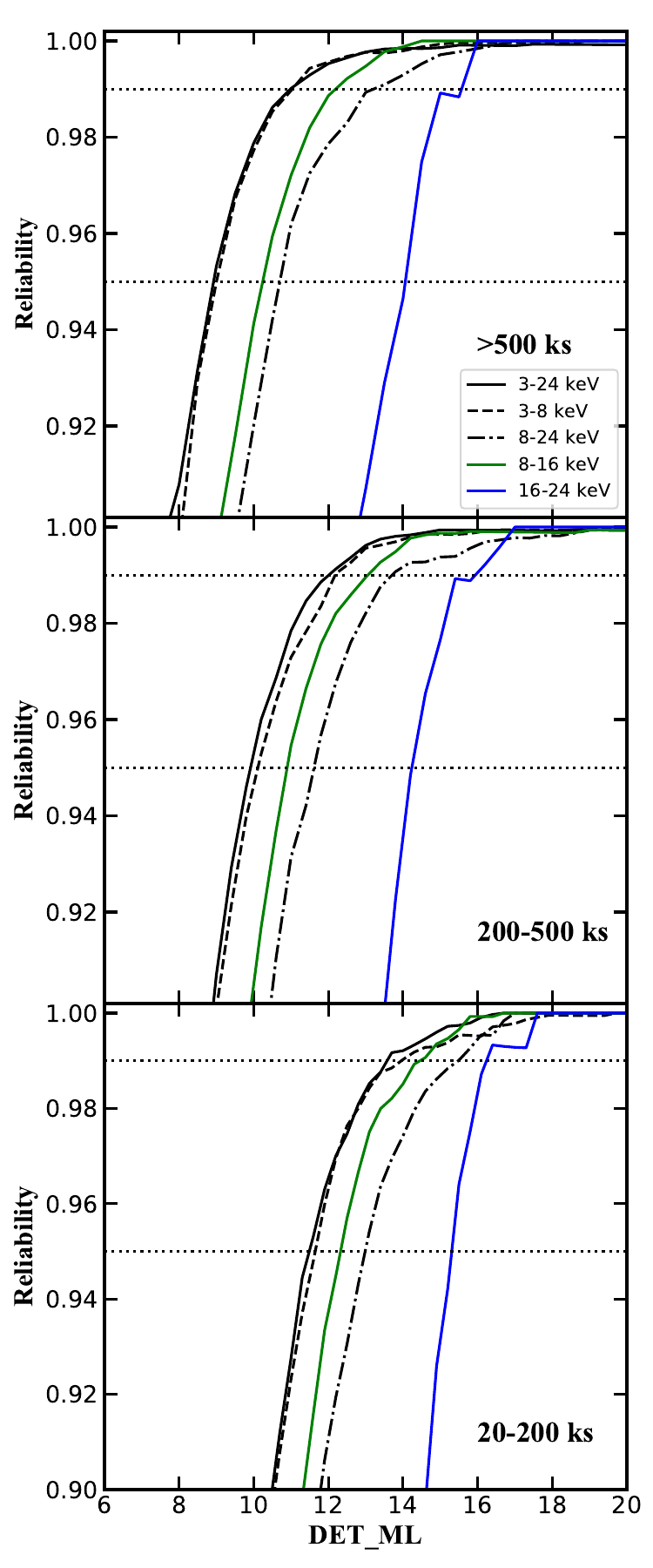}\par
\includegraphics[width=.5\textwidth]{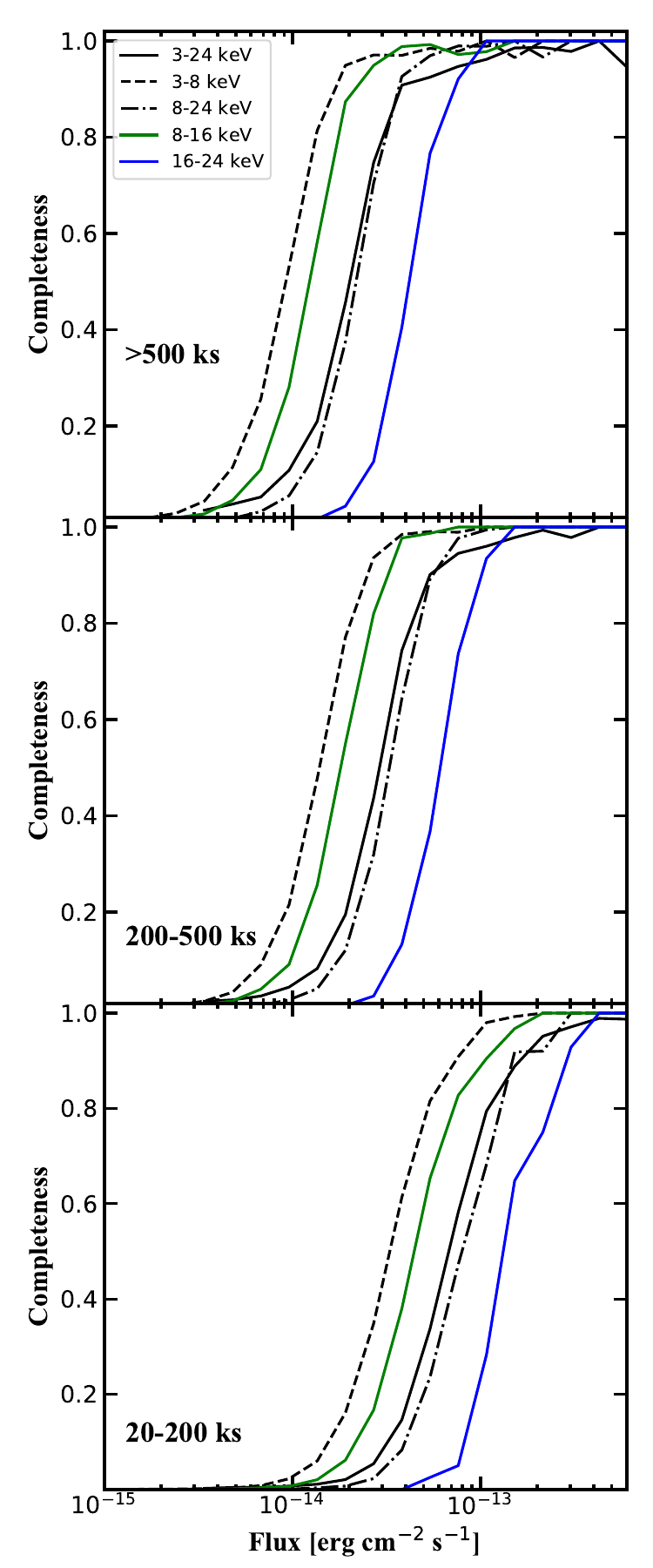}\par
\end{multicols}
\vspace{-0.3cm}
\caption{From top to bottom are cumulative reliability as a function of DET\underline{\;\;}ML (left) and completeness as a function of flux (right) in $>$500\,ks (top), 200-500\,ks (middle), and 20-200\,ks (bottom) exposure intervals. Solid black: 3--24\,keV. Dashed black: 3--8\,keV. Dash-dot black: 8--24\,keV. Sold green: 8--16\,keV. Solid blue: 16--24\,keV. The 99\% and 95\% reliability lines are plotted using dotted black lines on left.}
\label{fig:reliability}
\end{figure*}   

After performing the source detection, we obtain a catalog of detections for each simulation and each energy band. The average numbers of sources detected above the detection limit in each band are listed in Table~\ref{Table:number}. The counts associated with each detected source may be contaminated by the nearby sources within 90\arcsec\ due to the point-spread function. Therefore, we apply a deblending process to the detections in the catalogs of each simulation following the technique in \citet{Mullaney15}. The deblended source counts and background counts were then used to compute updated DET\underline{\;\;}ML values for each detection. The detections were matched with the input catalog using a 30\arcsec\ separation limit. The average numbers of matched sources for each simulation in five energy bands are reported in Table~\ref{Table:number}. 

\begin{figure} 
\centering
\includegraphics[width=.5\textwidth]{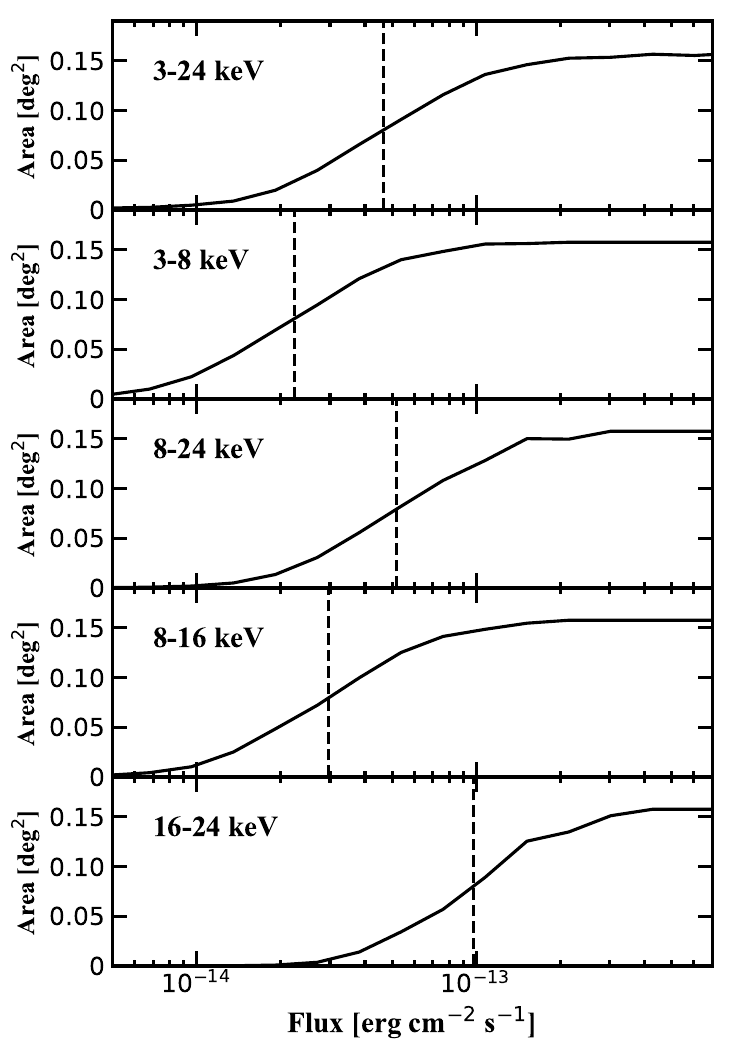}
\caption{Sky coverage in five energy bands at 95\% reliability level. The half-area flux is labeled for each energy band.}
\label{fig:coverage}
\end{figure}

\subsection{Reliability, Completeness, Positional Uncertainty} \label{sec:relia}
With extensive simulations, we calculate the reliability and completeness of our source detection technique. Reliability is the ratio of the number of the detected sources matched to input sources to the total number of detected sources at or above a particular DET\underline{\;\;}ML. For example, if 100 sources were detected above a certain DET\underline{\;\;}ML and 95 of them had counterparts from the input source catalog, then the reliability of the source detection at that DET ML is 95\%. Completeness is defined as the ratio of the number of the detected sources that are matched to the input catalog and are above a chosen reliability threshold to the number of the sources in the input catalog at a particular flux. For example, if 50 sources were detected and with DET\underline{\;\;}ML are above 95\% reliability threshold at a particular flux, then the completeness at that flux is 50\% at 95\% reliability level if there are 100 input sources at that flux. Sources with larger DET\underline{\;\;}ML are considered to be more reliable, but a higher DET\underline{\;\;}ML threshold leads to lower completeness. Therefore, to determine the detection criteria for the actual observations, we need to maximize both the survey completeness and reliability. 

We found that the ratio of the spuriously detected sources to all detected sources decreases exponentially as the exposure time increases. Thus, the reliability curve as a function of DET\underline{\;\;}ML is exposure-dependent as well. Due to the design of the \NuSTAR\ NEP survey, the exposure time across the field is nonuniform. We analyze the reliability function at three different exposure intervals (combined FPMA+FPMB), i.e., 20--200\,ks, 200--500\,ks, and $>$500\,ks. We applied an exposure cut-off at 20\,ks to avoid potentially spurious detections on the edge of the observations. The 200\,ks and 500\,ks values were chosen to evenly separate the total exposure $\sim$680\,ks (see Fig.~\ref{fig:expomap}). In Fig.~\ref{fig:reliability} left, we plot the reliability distribution as a function of DET\underline{\;\;}ML in different exposure intervals. The maximum likelihoods corresponding to 99\% and 95\% reliability (i.e., 1\% and 5\% spurious detections) in different exposure intervals are reported in Table~\ref{Table:number}, respectively. We found that the DET\underline{\;\;}ML thresholds are significantly different among the five energy bands and among different exposure intervals, which was also found in previous studies \citep[e.g.,][]{Civano_2015,Masini_2018}. The DET\underline{\;\;}ML thresholds of 3--24\,keV and 3--8\,keV are much lower than the DET\underline{\;\;}ML thresholds in hard energy bands, and the DET\underline{\;\;}ML thresholds decreases significantly as the exposure time increases. The average numbers of the sources which were detected above 99\% and 95\% reliability threshold in the simulated maps are reported in Table~\ref{Table:number}. 

\begin{figure} 
\centering
\includegraphics[width=.5\textwidth]{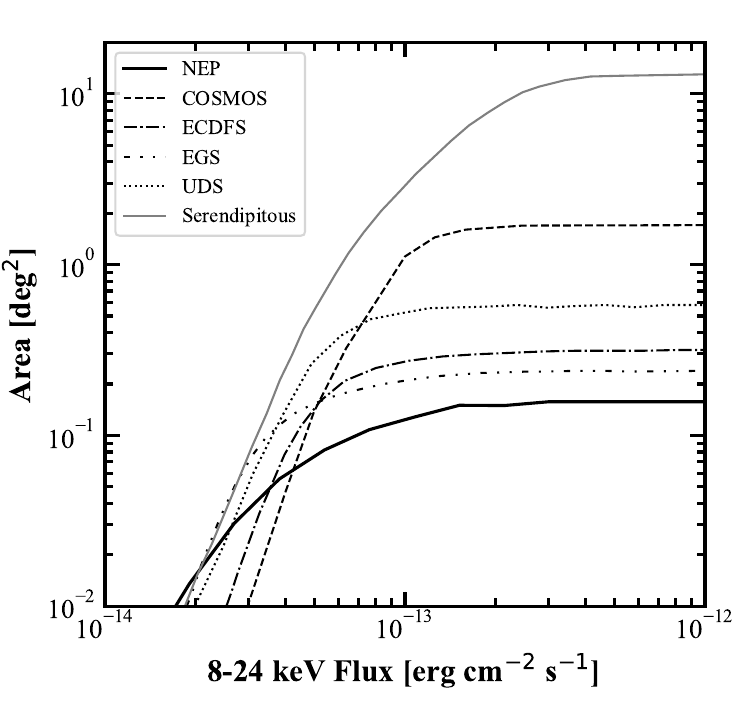}
\caption{Hard band (8--24\,keV) sky-coverage of \NuSTAR\ surveys: NEP in black solid line, COSMOS \citep{Civano_2015} in dashed line, ECDFS \citep{Mullaney15} in dashdot line, EGS (Aird et al., in prep) in dashdotdotted line, UDS \citep{Masini_2018} in dotted line, and Serendipitous survey \citep{Lansbury_2017} in gray solid line.}
\label{fig:sensitivity}
\end{figure}

After determining the DET\underline{\;\;}ML thresholds for different reliability levels, in Fig.~\ref{fig:reliability} right, we plot the detection completeness as a function of source flux of five energy bands at 95\% reliability level in different exposure intervals. The completeness curves for each energy band also depend on the exposure intervals, as the completeness value at a certain flux increases significantly as the exposure increases. The completeness is the fraction of the sources in the survey FoV that are detected and above the reliability threshold. Therefore, the completeness corresponds to the normalization of the effective sky-coverage of the survey which is the area covered at a given flux. The sky-coverage of the survey at a particular flux is thus the completeness of the survey at that particular flux multiplied by the maximum covered area exhibited in the exposure map. The total sky-coverage of the survey was calculated by adding up the sky-coverage curves of all three exposure intervals. The half-area fluxes are 4.67, 2.24, 5.19, 2.97, and 9.75 $\times$ 10$^{-14}$ erg\,cm$^{-2}$\,s$^{-1}$ for 3--24, 3--8, 8--24, 8--16, and 16--24\,keV, respectively. The fluxes at 20\% of the sky-coverage are 2.39, 1.14, 2.76, 1.52, and 5.20 $\times$ 10$^{-14}$\,erg\,cm$^{-2}$\,s$^{-1}$ in the 3--24, 3--8, 8--24, 8--16, and 16--24\,keV bands, respectively. The sky-coverage of the survey in the five energy bands is plotted in Fig.~\ref{fig:coverage}. The hard band (8--24\,keV) sky-coverage is plotted in Fig.~\ref{fig:sensitivity}, which is compared with the previous \NuSTAR\ surveys \citep[COSMOS;][]{Civano_2015}, \citep[ECDFS;][]{Mullaney15}, (EGS; Aird et al., in prep), and \citep[Serendipitous; ][]{Lansbury_2017}. Reaching the deepest flux in the hard X-ray band makes NEP one of the most sensitive \NuSTAR\ contiguous surveys to date.

\begin{figure} 
\centering
\includegraphics[width=.48\textwidth]{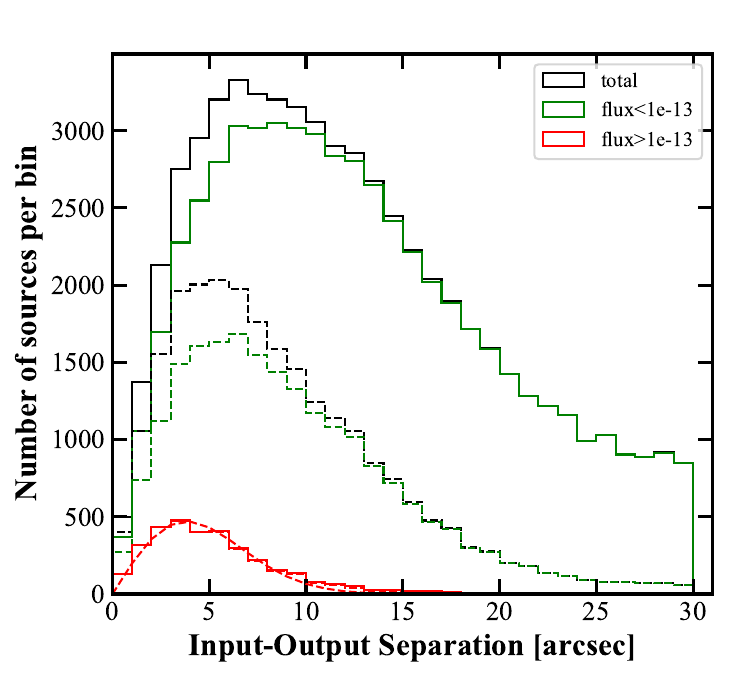}
\caption{Distributions of input to detected positions in the 3--24\,keV bands for the 1200 simulations. Solid refer to the whole sample and dashed lines refer only to sources above the 95\% reliability DET\underline{\;\;}ML thresholds. The black lines represent the distribution of total sources, the green lines represent the distribution for sources with 3--24 keV flux $<$10$^{-13}$ erg\,cm$^{-2}$s$^{-1}$, and the red lines represent sources with brighter fluxes. The dashed red curve is the best Rayleigh-fit of the input-output separation of the bright sources above the 95\% reliability DET\underline{\;\;}ML threshold.}
\label{fig:separation}
\end{figure}

In order to quantify the positional uncertainties of our detections, we plot the distribution of the separation between positions of the input and output detected sources in the 3--24\,keV band in Fig.~\ref{fig:separation} using solid lines. The separation distribution follows a Rayleigh distribution \citep{Pineau17}, thus we performed a Rayleigh fitting of the detected separation distribution. We found a best-fit Rayleigh scale parameter of $\sigma_{all}$ = \ang[angle-symbol-over-decimal]{;;9.7} for all matched sources. The separations between the input and output detected positions are significantly improved when only using the sources with DET\underline{\;\;}ML above 95\% reliability threshold ($\sigma_{95\%}$ = \ang[angle-symbol-over-decimal]{;;6.3}). The input-output separations are smaller for brighter sources with 3--24\,keV flux $>10^{-13}$ erg\,cm$^{-2}$s$^{-1}$ ($\sigma_{95\%,bright}$ = \ang[angle-symbol-over-decimal]{;;3.8}) than the fainter sources ($\sigma_{95\%,faint}$ = \ang[angle-symbol-over-decimal]{;;6.9}). The fitted separations are consistent with previous work. Therefore, we can adopt their distribution as the expected positional uncertainty of true detections.

\subsection{Fluxes}
\begin{figure} 
\centering
\includegraphics[width=.49\textwidth]{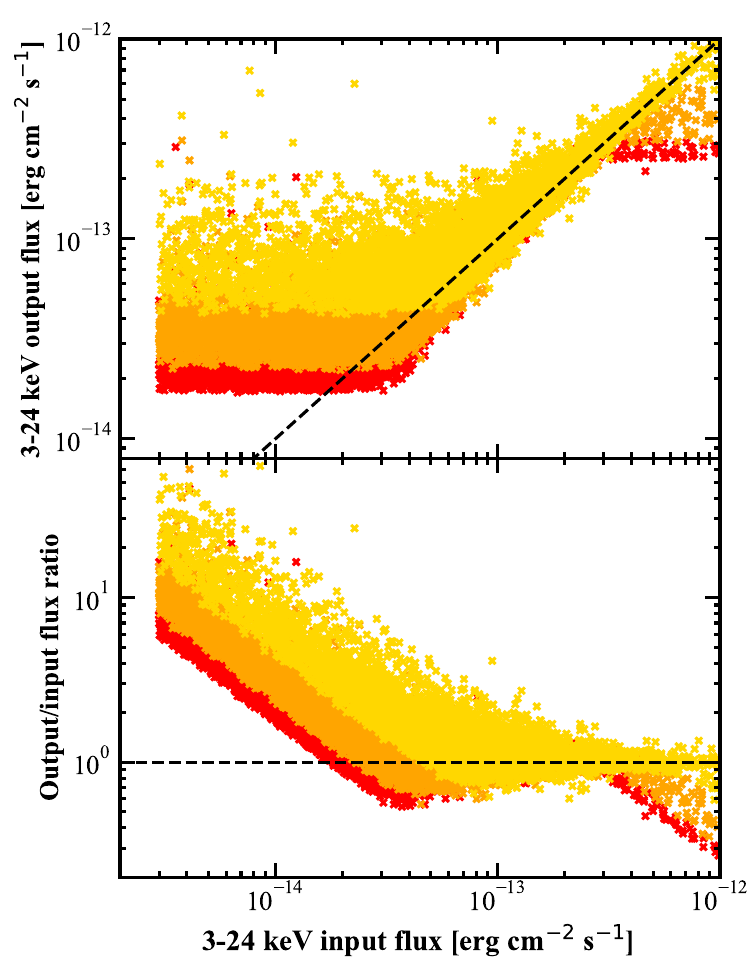}
\caption{Upper: 3--24 keV input fluxes versus the measured fluxes of the sources with DET\underline{\;\;}ML above the 95\% reliability threshold in the simulations with exposure of $>$500\,ks (red), 200--500\,ks (orange), and 20--200\,ks (yellow). Lower: Ratio between measured fluxes and input fluxes as a function of input fluxes.}
\label{fig:flux}
\end{figure}

The flux of each matched source in the simulations was computed in different energy bands. We extracted the source counts and deblended background counts of each matched source using the CIAO \citep{CIAO} tool {\tt dmextract}. The net counts were then converted to fluxes using the exposure time at the source position and the count-rate-to-flux CF mentioned in Section~\ref{sec:simulation}. Since the counts were extracted in a 20\arcsec\ area, we convert this aperture flux to the total flux using a factor, i.e., $F_{20\arcsec}/F_{tot}$ = 0.32, calculated from the \NuSTAR\ PSF\footnote{\url{https://heasarc.gsfc.nasa.gov/docs/nustar/NuSTAR_observatory_guide-v1.0.pdf}}. We used a single $F_{20\arcsec}/F_{tot}$ for FPMA and FPMB and for different energy bands as we found them to be consistent. In Fig.~\ref{fig:flux}, we plot the 3--24\,keV input fluxes ($F_{in}$) and the measured 3--24\,keV fluxes of the corresponding matched sources with DET\underline{\;\;}ML above 95\% reliability threshold in different exposure intervals. The flux limits of the detections are $\sim$1.7 $\times$ 10$^{-14}$\,erg\,cm$^{-2}$\,s$^{-1}$ for $>$500\,ks exposure interval, $\sim$2.2 $\times$ 10$^{-14}$\,erg\,cm$^{-2}$\,s$^{-1}$ for 200--500\,ks exposure interval, and $\sim$4.2 $\times$ 10$^{-14}$\,erg\,cm$^{-2}$\,s$^{-1}$ for 20--200\,ks exposure interval, which are consistent with the flux limits of the completeness curves shown in the right panel of Fig.~\ref{fig:reliability}. As seen in Fig. ~\ref{fig:flux}, at faint fluxes ($F_{in}$ $<$5 $\times$ 10$^{-14}$\,erg\,cm$^{-2}$\,s$^{-1}$) that approach the detection limit, the measured flux distribution widens due to the Eddington bias. At bright fluxes ($F_{in}$ $>$3 $\times$ 10$^{-13}$\,erg\,cm$^{-2}$\,s$^{-1}$), the measured values might be underestimated due to the fixed aperture used to extract the \NuSTAR\ counts. This underestimated effect does not impact the flux measurement of the bright sources in the NEP survey, since the 3--24\,keV flux of the brightest source is 2.8 $\times$ 10$^{-13}$\,erg\,cm$^{-2}$\,s$^{-1}$ (see Section~\ref{sec:catalog}).

\section{Source Catalog} \label{sec:catalog}
\begin{figure*} 
\begin{multicols}{2}
\includegraphics[width=.5\textwidth]{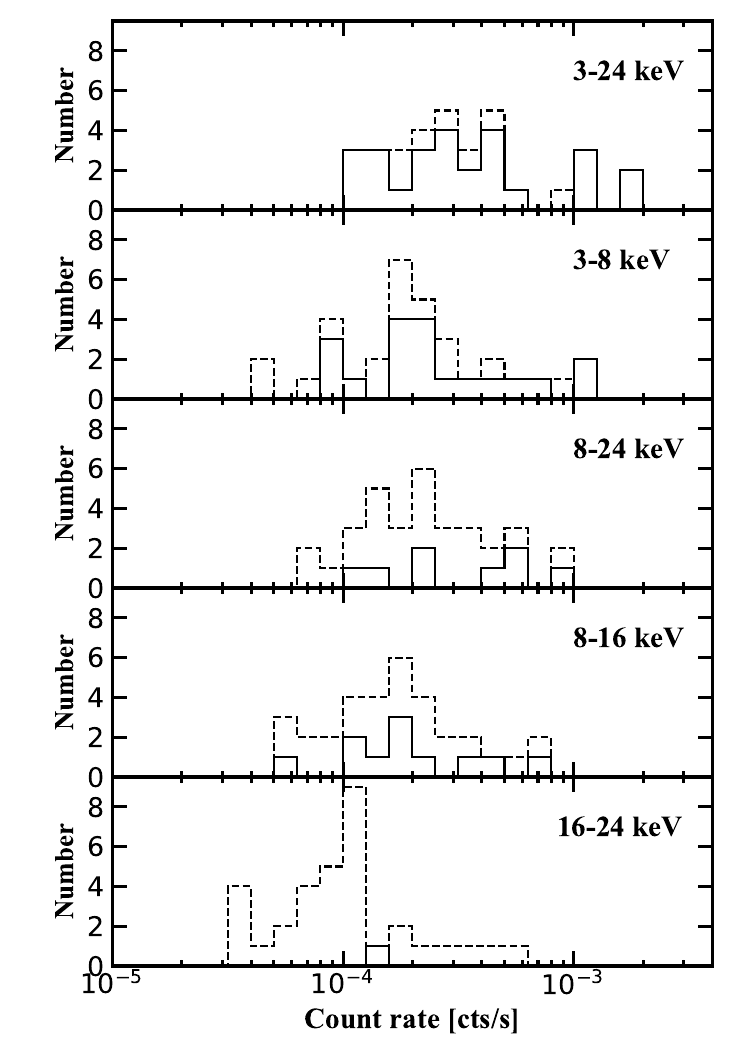}\par
\includegraphics[width=.5\textwidth]{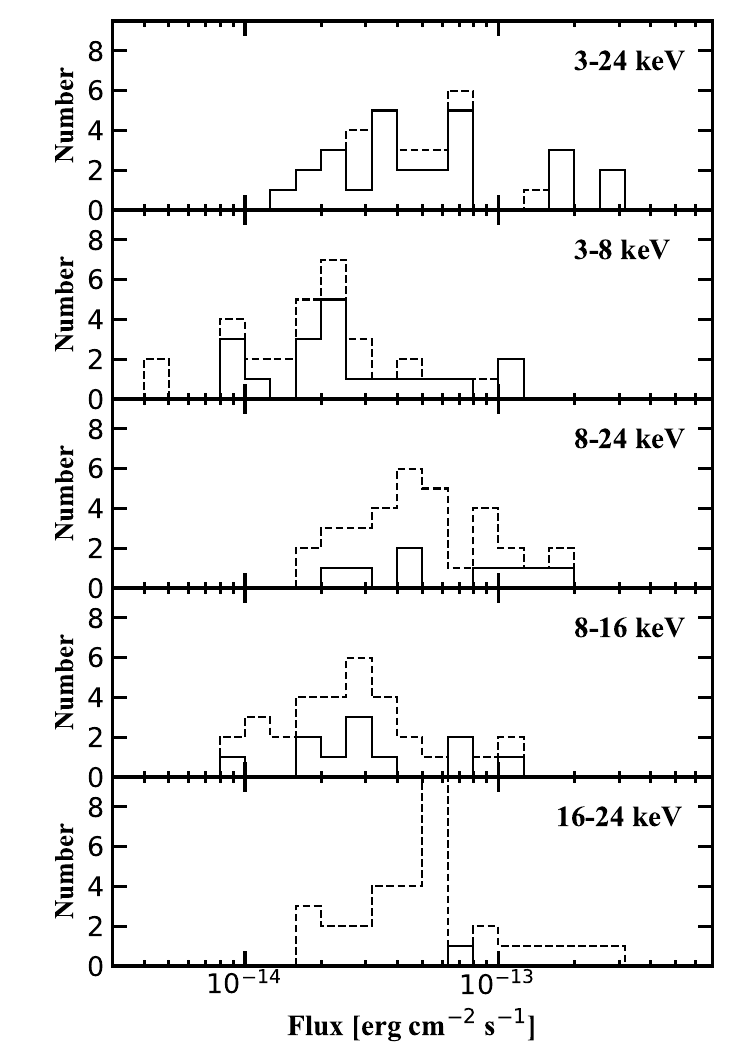}\par
\end{multicols}
\vspace{-0.3cm}
\caption{20\arcsec\ radius aperture count rate (left) and total flux (right) distributions for sources detected and above threshold (solid line) in the 3--24, 3--8, 8--24, 8--16, and 16--24\,keV bands from top to bottom. Upper limits are plotted in dashed lines.}
\label{fig:cts_flux}
\end{figure*}  

We performed source detection on the observed data using the same technique as described in Section~\ref{sec:detection} by using \textit{SExtractor}. We extracted the total counts and the deblended background counts of the detected sources on the FPMA+FPMB mosaics and the simulated background maps using a 20\arcsec\ radius aperture. Using the extracted counts, we calculated DET\underline{\;\;}ML of each detection in different energy bands. Thus we generated a catalog of sources having DET\underline{\;\;}ML above the 99\% and 95\% reliability thresholds for each energy band. Then we merged the catalog of each band to make a master catalog by performing a positional matching (30\arcsec\ in radius) among the catalogs of five energy bands. The master catalog consists of the sources that are detected and pass the DET\underline{\;\;}ML thresholds in at least one energy band (including both matched and unmatched sources). It is worth noting that the DET\underline{\;\;}ML thresholds depend on the source exposure. The master catalog contains 21 sources above the 99\% reliability threshold and 33 sources above the 95\% reliability threshold. In the following discussion, we focus on the 95\% reliability catalog. Statistically, we expected $\sim$2 spurious detections in the 95\% reliability catalog (summing the expected number of spuriously detected sources in each band). 
We report the distribution of the detected sources in the master catalog in Table~\ref{Table:FSH}, including sources that are detected and pass above the 95\% reliability threshold labeled with capital F, S, and H for 3--24\,keV, 3--8\,keV and 8--24\,keV (and/or 8--16\,keV and 16--24\,keV), and sources that are detected but fall below the threshold labeled with lower case f, s, and h in given energy bands.

The coordinates of the sources listed in the final catalog are taken from the energy band corresponding to the highest DET\underline{\;\;}ML. The position of each detected source is plotted in Fig.~\ref{fig:mosaic}. The vignetting-corrected exposure times at the positions of the sources are measured in each energy band. The total counts and background counts (deblended if detected and passing the DET\underline{\;\;}ML threshold) of each source in a given energy band are measured at the positions of the sources in a 20\arcsec\ radius aperture. The net counts (total counts -- deblended background counts) and 1\,$\sigma$ net counts errors were calculated for the sources that were detected and passed the DET\underline{\;\;}ML threshold in a given energy band. The 90\% confidence upper limit of the net counts was calculated for undetected sources, or the sources that were detected but fell below the DET\underline{\;\;}ML threshold in a given energy band. The net counts errors and upper limits were determined using Equations (9) and (12) in \citet{Gehrels1986} with $S$ = 1 for 1\,$\sigma$ confidence level and $S$ = 1.645 for 90\% confidence level. The count rates are calculated by dividing the net counts by the vignetting-corrected exposure time in a given energy band. The total fluxes in a given energy band were converted from the count rates using the count-rate-to-flux CF described in Section~\ref{sec:simulation} and were corrected the aperture correction factor ($F_{tot}$/$F_{20\arcsec}$ $\sim$0.32 for 20\arcsec\ aperture). In Fig.~\ref{fig:cts_flux}, the distribution of the 20\arcsec\ radius aperture count rates and total fluxes in different energy bands of all sources in the catalog are plotted (90\% confidence upper limits are also included).

We also performed source detection on the data in 35--55\,keV band, where the background can be well characterized using {\tt nuskybgd}. By using the same source detection and deblending technique, we computed the DET\underline{\;\;}ML for each detection in \textit{SExtractor}. We adopted the DET\underline{\;\;}ML threshold of the 35--55\,keV band computed in \citet{Masini_2018}, which used the same simulation analysis as we did for other energy bands. We did not find any significant detection that was above the 95\% reliability threshold (DET\underline{\;\;}ML = 22.2). The most significant detection in 35--55\,keV band is DET\underline{\;\;}ML = 10.75. The non-detection in 35--55\,keV band is consistent with \citet{Masini_2018}, who did not detect any source in 35--55\,keV band in the $\sim$0.6\,deg$^2$ UDS field.

\begingroup
\renewcommand*{\arraystretch}{1.2}
\begin{table}
\centering
\caption{Distribution of the detected sources in the 95\% reliability catalog. F(f), S(s), and H(h) represent 3--24\,keV, 3--8\,keV, 8--24\,keV (and/or 8--16\,keV and 16--24\,keV). F, S, H represent sources detected and above the 95\% reliability threshold in the given energy band, while f, s, and h refer to the sources detected but below the 95\% reliability threshold.}
\label{Table:FSH}
  \begin{tabular}{lcr}
       \hline
       \hline     		F+S+H&\quad\quad\quad\quad\quad\quad\quad\quad\quad\quad\quad\quad\quad\quad\quad\quad&9 (27\%)\\
	F+S+h&&6 (18\%)\\
	F+S&&1 (3\%)\\
	F+s+h&&5 (15\%)\\
	F+s&&3 (9\%)\\
	F+h&&2 (6\%)\\
	f+S+h&&1 (3\%)\\
	f+S&&1 (3\%)\\
	S&&1 (3\%)\\
	f+s+H&&1 (3\%)\\
	f+H&&2 (6\%)\\
	H&&1 (3\%)\\
	\hline
	\hline
\end{tabular}
\end{table}
\endgroup

\subsection{log$N$-log$S$}
\begin{figure} 
\centering
\includegraphics[width=.49\textwidth]{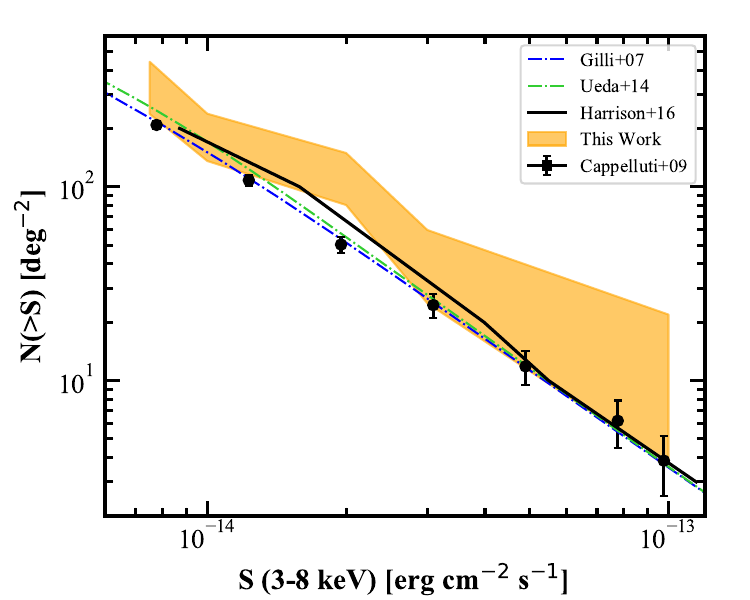}
\includegraphics[width=.49\textwidth]{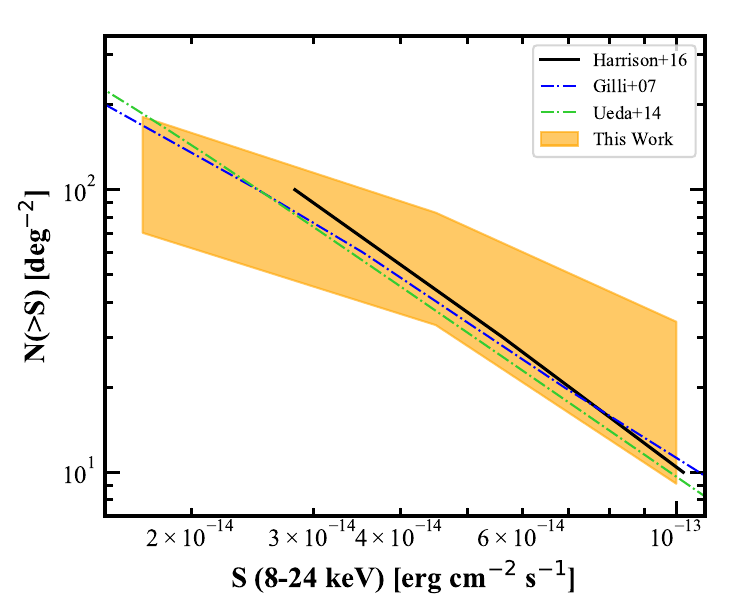}
\includegraphics[width=.49\textwidth]{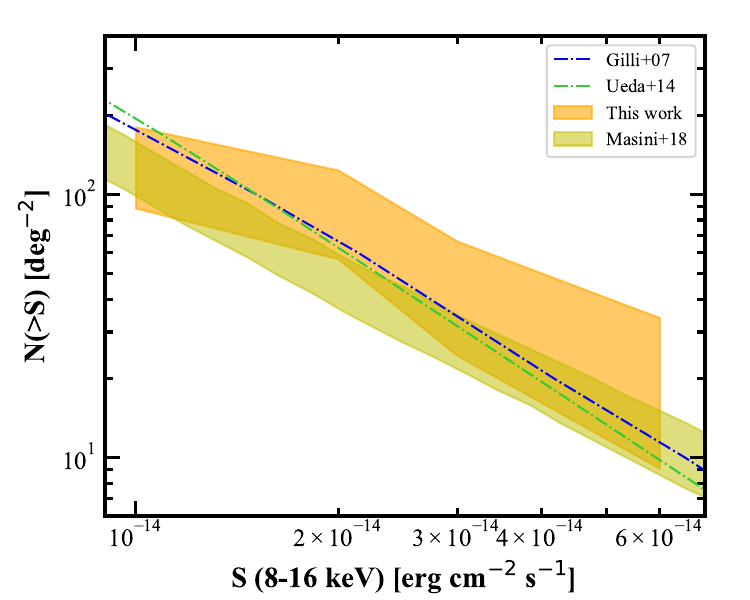}
\vspace{-0.2cm}
\caption{Integral number counts for the 3--8\,keV (top), 8--24\,keV (middle), and 8--16\,keV (bottom) bands. The orange shaded region represents the 68\% confidence region on the integrated number counts. The results of our work are compared with those derived in \citet{harrison15} (black solid line) and \citet{Masini_2018b} (yellow shaded region) using \NuSTAR, those derived in \citet{Cappelluti09} (black points) using \XMM, and those constrained by the population synthesis model of CXB \citep[blue dashdot line,][]{gilli07} and \citep[green dashdot line,][]{Ueda14}.}
\label{fig:logN-logS}
\end{figure}

We computed the cumulative log$N$-log$S$ distribution in three energy bands (3--8\,keV, 8--24\,keV, and 8--16\,keV) using the sky coverage as a function of flux following \citet{Cappelluti09}. The distribution is defined as
\begin{equation}
N(>S)=\sum_{i=1}^{N_S}\frac{1}{\Omega_i} \rm deg^{-2},
\end{equation}
where $N(>S)$ is the total number of sources detected and passing the 95\% reliability threshold in that energy band with flux greater than $S$. $\Omega_i$ is the sky coverage associated with the flux of the ith source. The variance of the number counts is thus defined as 
\begin{equation}
\sigma^2_S = \sum_{i=1}^{N_S}(\frac{1}{\Omega_i})^2.
\end{equation}
For the minimum flux, we use as a factor of three lower than the flux corresponding to 50\% of the survey sky-coverage in each band. The maximum flux is 10$^{-13}$\,erg\,cm$^{-2}$\,s$^{-1}$ for 3the --8\,keV and 8--24\,keV bands, and 6 $\times$ 10$^{-14}$\,erg\,cm$^{-2}$\,s$^{-1}$ for the 8--16\,keV band. The cumulative log$N$-log$S$ in the 3--8\,keV, 8--24\,keV, and 8--16\,keV bands are plotted in Fig.~\ref{fig:logN-logS}. Our work extends the 8--24\,keV number counts to lower fluxes than previous work presented in \citet{harrison15}. The soft band number counts are compared with those derived using \XMM\ by \citet{Cappelluti09}. The 2--10\,keV fluxes reported in \citet{Cappelluti09} are converted to 3--8\,keV assuming a power law model with photon index $\Gamma$ = 1.80. The log$N$-log$S$ constrained by the CXB population synthesis models, e.g., \citet{gilli07,Ueda14} are shown as points of comparison as well. The 3--8\,keV number counts in our work surpass those measured or constrained in the previous observations and models, but are still within the 1$\sigma$ uncertainties. Our cumulative log$N$-log$S$ in hard energy bands, i.e., 8--24\,keV and 8--16\,keV, are in good agreement with the previous results derived from \NuSTAR\ \citep{harrison15,Masini_2018b} and with the CXB population synthesis models.

\section{Multiwavelength counterparts}\label{sec:multi}
\subsection{Soft X-ray Counterparts}
In order to associate the \NuSTAR\ detected sources with the multiwavelength counterparts, accurate coordinates are needed. The best strategy is first to identify the soft X-ray counterpart of the \NuSTAR\ detected sources, which typically have more accurately measured positions. Currently, there is no public soft X-ray (e.g., \cha\ and \XMM) catalog in the {\it JWST} NEP Time-Domain Field.

\subsubsection{\XMM}
In \NuSTAR\ cycle 6, we were granted additional \NuSTAR\ observations together with 40\,ks \XMM\ observations (PI: Civano) to study hard X-ray variability with simultaneous broad X-ray coverage in the 0.5--24\,keV band in the NEP field. The first 10\,ks \XMM\ observation was taken on Oct. 14th, 2020. The \XMM\ observation covered about 90\% of the \NuSTAR\ NEP field with the exception of the two bottom corners of the field (where only one \NuSTAR\ source, ID 26, was detected).

In order to obtain a list of soft X-ray sources in the \NuSTAR\ NEP field, we performed source detection on the first 10\,ks \XMM\ observation. We used similar source detection techniques as in the previous \XMM\ surveys \citep[e.g.,][]{Brunner08,Cappelluti09,LaMassa_2016}, which are briefly summarized here. The observational data files (ODF) of MOS1, MOS2, and pn were generated using the SAS (version 19.0.0) tasks {\tt emproc} and {\tt epproc}. The background flares of the three instruments were screened using the standard method. Since the \XMM\ background was contaminated by the Al k$\alpha$ fluorescent emission at 1.48\,keV in both MOS and pn data and by two Cu lines at $\sim$7.4\,keV and 8.0\,keV in pn data, the events between 1.45\,keV and 1.54\,keV were removed in both MOS and pn data, and the events in 7.2--7.6\,keV and 7.8--8.2\,keV were also removed from the pn data. 

The images of the clean event files of MOS1, MOS2, and pn were then generated in the 0.5--2\,keV and 2--10\,keV energy bands. To fully utilize the sensitivity of the \XMM\ data, we created the MOS+pn mosaics in each energy band using the SAS {\tt emosaic} task. The exposure maps were generated using the SAS task {\tt eexpmap} for each instrument in the two energy bands. The energy conversion factors (ECFs), used to convert from count rates to flux, were calculated using WebPIMMs for each energy band and each instrument, assuming an absorbed power-law model with $\Gamma$ = 1.80 and Galactic column density of N$\rm _H$ = 3.4 $\times$ 10$^{20}$ cm$^{-2}$. The ECFs for MOS and pn are 0.54 and 0.15 $\times$ 10$^{-11}$ erg\,cm$^{-2}$\,counts$^{-1}$ in 0.5--2\,keV, and 2.22 and 0.85 $\times$ 10$^{-11}$ erg\,cm$^{-2}$\,counts$^{-1}$ in 2--10\,keV, respectively. For each of the two energy bands, the exposure maps of the three instruments were co-added, weighted by their ECFs to create mosaics using the HEASoft {\tt fcarith} tool and the SAS {\tt emosaic} task. The background maps were simulated after masking out the bright sources (with detection likelihood {\tt LIKE}\footnote{{\tt LIKE} = --~ln~$p$, where $p$ is the probability of Poissonian random fluctuation of the counts in the detection cell which would have resulted in at least the observed number of source counts.}$>$4). The background map mosaics were then generated in the two energy bands. The source detection was performed on the mosaics of the three instruments using the SAS {\tt eboxdetect} (with {\tt LIKE}$>$4) and {\tt emldetect} (with {\tt LIKE}$>$6) tasks. The reliabilities of the chosen likelihood level correspond to 97.3\% in 0.5--2\,keV and 99.5\% in 2--10\,keV respectively according to the simulations of \citet{Cappelluti_2007}. The sources were detected simultaneously in the 0.5--2\,keV and 2--10\,keV energy bands, thus our catalog contains sources with {\tt LIKE}$>$6 ($\ge3\,\sigma$) in either 0.5--2\,keV or 2--10\,keV band, or the sources with the equivalent 0.5--10\,keV detection likelihood\footnote{The {\tt LIKE} values from each individual energy band are added and transformed to equivalent single band detection likelihoods using the incomplete Gamma function. Refer to \url{https://xmm-tools.cosmos.esa.int/external/sas/current/doc/eboxdetect/node3.html} for further details} {\tt LIKE}$>$6. Our catalog includes 165 sources: 88 sources with {\tt LIKE}$>$6 in 0.5--2\,keV band and 71 sources with {\tt LIKE}$>$6 in 2--10\,keV. The 2--10\,keV flux lower limit of the sources with {\tt LIKE}$>$6 is F$\rm _{2-10\,keV}$ $\approx$ 8 $\times$ 10$^{-15}$ erg\,cm$^{-2}$\,s$^{-1}$.

\begin{figure} 
\centering
\includegraphics[width=.48\textwidth]{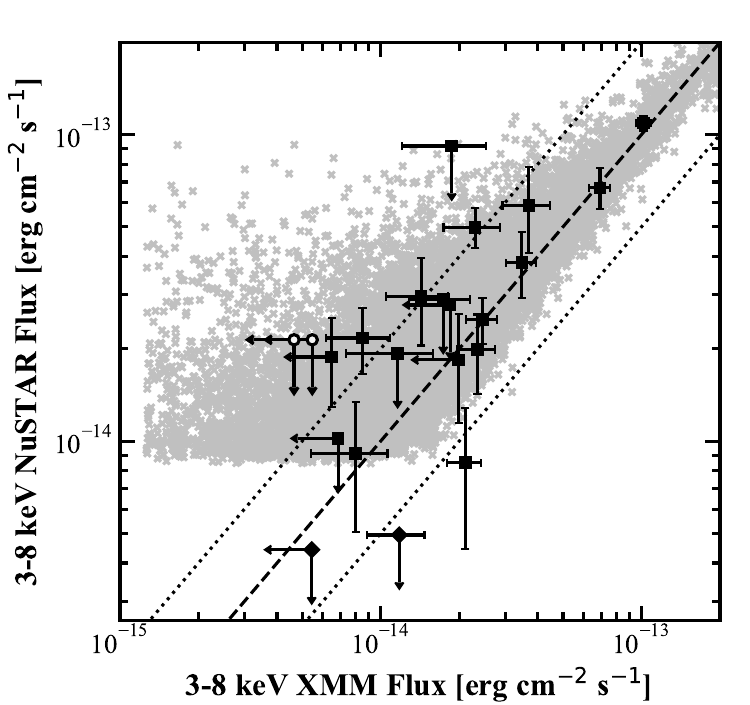}
\caption{Comparison between 3--8\,keV fluxes measured in \NuSTAR\ and in \XMM. The dashed line represents the 1:1 relation, while the dotted lines are a factor of two from the 1:1 relation. ID 30 and 33 and are plotted as the black diamonds. The black open circles are the two \XMM\ counterparts of \NuSTAR\ ID 31. The gray crosses on the background are the input and output fluxes from the simulations of the sources detected and above the 95\% reliability threshold} in the 3--8\,keV band. The 3--8\,keV \NuSTAR\ fluxes of sources that are not detected in S band are plotted as upper limits.
\label{fig:flux_compare}
\end{figure}

We cross-matched the \NuSTAR\ catalog with the \XMM\ catalog by using a radius combining a 20\arcsec\ matching radius and the positional error of \XMM\ read from the {\tt emldetect} best-fit results in quadrature (a 1\arcsec\ systematic error was added to the \XMM\ positional error in quadrature). The 20\arcsec\ matching radius is chosen to be three times of the best-fit Rayleigh scale parameter of the input-output separations ($\sigma_{95\%}$ = \ang[angle-symbol-over-decimal]{;;6.3}) from the simulations (see, Fig.~\ref{fig:separation}) to ensure that all potentially faint sources can be matched. 19 \NuSTAR\ detected sources have \XMM\ associations within the matching radius. \NuSTAR\ source ID 31 has double \XMM\ counterparts located \ang[angle-symbol-over-decimal]{;;16.1} and \ang[angle-symbol-over-decimal]{;;18.9} away from the \NuSTAR\ detected position, although only one of these sources is detected with \cha\ (see Section~\ref{sec:cha}). The two \XMM\ sources are in the opposite direction of the \NuSTAR\ detected position. Therefore, \NuSTAR\ might detect the combination of two soft X-ray sources. 

One \XMM\ detection (with 2--10\,keV $likemin$ = 13.7) is \ang[angle-symbol-over-decimal]{;;21.5} away from \NuSTAR\ source ID 6. The \NuSTAR\ 3--24\,keV flux of ID 6 is $F_{3-24}$ = 5.5$\pm$1.1 $\times$ 10$^{-14}$ erg\,cm$^{-2}$\,s$^{-1}$. Another \XMM\ detection (with 2--10\,keV $likemin$ = 27.1) is \ang[angle-symbol-over-decimal]{;;23.7} away from \NuSTAR\ source ID 27. The \NuSTAR\ 3--24\,keV flux of ID 27 is $F_{3-24}<$ 1.4 $\times$ 10$^{-14}$ erg\,cm$^{-2}$\,s$^{-1}$. The simulations of the input-output separation of faint sources with flux $F_{3-24}<$ 10$^{-13}$ erg\,cm$^{-2}$\,s$^{-1}$ suggest that 6.5\% out of the 28 detected faint sources, which is $\sim$2 sources, might have a $>$20\arcsec\ distance between the \NuSTAR\ and the soft X-ray detected coordinates (see, Fig.~\ref{fig:separation}). Further more, the 3--8\,keV \NuSTAR\ fluxes of ID 6 and ID 27 and the 3--8\,keV flux of the two aforementioned \XMM\ detections are consistent with the simulations of input-output fluxes distribution as shown in Fig.~\ref{fig:flux_compare}. Therefore, we associate the two \XMM\ detections as the soft X-ray counterpart of ID 6 and ID 27. Thus, a total of 21 \NuSTAR\ sources have detected \XMM\ counterparts.

The 2--10\,keV fluxes and flux errors (1$\sigma$) of the \XMM\ detections were computed using the {\tt emldetect} task if the 2--10\,keV maximum likelihood of the sources is DET\underline{\;\;}ML $>$6. For sources whose 2--10\,keV maximum likelihood is DET\underline{\;\;}ML $<$6, we report the 90\% upper limit of the flux using the same equation as described in Section~\ref{sec:catalog}. We plot the 3--8\,keV \NuSTAR\ and \XMM\ fluxes of the matched sources in Fig.~\ref{fig:flux_compare}. \NuSTAR\ ID 31 has two potential \XMM\ counterparts, so we plotted the fluxes of both \XMM\ counterparts. The relations between the \NuSTAR\ and \XMM\ fluxes of most sources are consistent with the expectation from the simulations (see, Fig.~\ref{fig:flux}). The \NuSTAR\ 3--8\,keV flux of ID 30 and 33 are below the simulations because they are only detected above the threshold in the hard band. Nevertheless, their 3--8\,keV flux upper limits are consistent with the \XMM\ ones within a factor of two. 

The \XMM\ source number density at the flux limit of the \XMM\ observation (F$\rm _{3-8\,keV}$ $\sim$ 5 $\times$ 10$^{-15}$ erg\,cm$^{-2}$\,s$^{-1}$) is 390\,deg$^{-2}$. The number of \XMM\ sources found by chance within the search area (with a radius of 20\arcsec) around each \NuSTAR\ source is $<$0.09. Therefore, all the matches are very likely to be real associations. We investigate the \NuSTAR\ detected sources that do not have \XMM\ cross-matched associations. Two sources (ID 26 and 28) are either out of the FoV of \XMM\ or on the gaps between the detectors of \XMM. SDSS J172421.74+654847.5 ($z$ = 0.833 measured by Double Spectrograph (DBSP) of the Palomar Observatory, ATel \# 12049) was a variable AGN discovered in the \cha\ observations in 2018 (ATel \# 11906), which is \ang[angle-symbol-over-decimal]{;;27.1} away from \NuSTAR\ ID 16. We found that ID 16 is also a variable source and its 3--8\,keV \NuSTAR\ flux in epoch one was consistent with the \cha\ measured flux in July 2018. We thus associate this \cha\ detected transient to \NuSTAR\ ID 16. ID 16 was not detected by \XMM\ in 2020 October as we expected it to be too faint (see its flux variabilities as a function of time in the upper panel of Fig.~\ref{fig:ID16}). The details of the flux variability of this source are discussed in Section~\ref{sec:var}. 

ID 3, 5, 10, 15, 19, and 20 were detected in 3--8\,keV band of \NuSTAR\ with reliability of 79\%, 78\%, 99.5\%, 99.96\%, 96.4\% and 89\%, but were not detected by \XMM. These sources could be variable sources, whose flux decreased significantly in less than a year \citep[sources with significant variability in fluxes were found in days to months in previous works, e.g.,][]{Risaliti_2002,Markowitz14}. They could also be spuriously detected sources by \NuSTAR, although we note that only $\sim$2 spuriously detected sources are expected in the \NuSTAR\ 95\% reliability catalog. ID 23, 25, and 32 were neither detected in the soft band by \NuSTAR\ nor by \XMM. Therefore, they might be nearby heavily obscured sources whose soft X-rays are significantly suppressed due to large obscuration, which is supported by their high hardness ratio computed in Section~\ref{sec:HR}. Since they do not have measured redshifts, we cannot exclude the possibility that they are spurious detections. Another possibility of the non-detection of these sources is that their soft X-ray fluxes were below the detection capability of the 10\,ks \XMM\ observation. The forthcoming \NuSTAR\ and \XMM\ observations will help better locate the positions of these sources, which may result in more matches, and distinguish from the different scenarios.

\subsubsection{\cha}\label{sec:cha}
To obtain better positional accuracy of the \NuSTAR\ detected sources, we also perform source detection using the \cha\ data. We selected the \cha\ NEP observation, which was taken close to the \XMM\ observation mentioned above to minimize the effect of variability. Therefore, a \cha\ observation on September 21st, 2020 (ObsID: 21651) with exposure of $\sim$20\,ks was used. The \cha\ observation covered an area of 17\arcmin\ $\times$ 17\arcmin, only $\approx$50\% of the total \NuSTAR\ NEP field. The \cha\ data were reduced using {\tt CIAO} software package \citep{CIAO} version 4.12 and \cha\ CALDB version 4.12.3. The source detection was performed using the {\tt wavdetect} algorithm in the 0.5--7\,keV band. The source image, exposure map, and PSF map were created by running the {\tt fluximage} script. The images were binned with a pixel size of 4\arcsec\ on a side. The {\tt wavdetect} algorithm was then run on each on-axis chip with scales of 1 and 2 pixels, assuming a false-positive probability threshold of 10$^{-6}$, equivalent to one possibly spurious detection in the FoV. 

Thirty-nine sources were detected in \cha, and 16 were matched with \NuSTAR\ sources using a 20\arcsec\ search radius. All 16 sources have also been detected by \XMM, although only one of the two \XMM\ counterparts of ID 31 was detected by \cha. Of the five sources which were detected by \XMM\ but not by \cha, ID 1, 24, 27, and 29 were out of the \cha\ FoV, and ID 2 was weakly detected by \cha\ with a signal-to-noise ratio of SNR = 3.55 at the coordinate of \XMM\ detection. 
The two sources (ID 26 and 28) which were out of the FoV or on the detecter gaps of \XMM\ were out of the FoV of \cha\ as well. ID 16 was neither detected by \cha\ in September 2020. Complete \cha\ associations of the \NuSTAR\ detected sources will be built after the release of the \cha\ NEP source catalog (Maksym et al, in prep), which will cover the entire \NuSTAR\ area.

Therefore, a total of 22 \NuSTAR\ sources have a soft X-ray counterpart from \XMM\ and/or \cha. The soft X-ray (\cha\ if having \cha\ counterpart, otherwise \XMM) coordinates of the associated source are listed in Table~\ref{Table:catalog}. The positions of the soft X-ray counterparts of each \NuSTAR\ sources are plotted in Fig.~\ref{fig:mosaic}. 

\subsection{Optical counterparts}
The {\it JWST} NEP Time-Domain Field has extensive multiwavelength coverage. We used two optical catalogs to match the \NuSTAR\ detected sources: SDSS DR16 \citep{Ahumada20} and {\it Subaru} Hyper Suprime-Cam (HSC; observed as part of the HEROES survey, PI: Hasinger; image reduced by S. Kikuta, catalog produced by C. Willmer using \textit{SExtractor}). We applied a magnitude cut to the HSC catalog by removing the detections with $i$-band magnitude (all magnitudes below are quoted on AB system) $m_i<$19 since bright sources are saturated in HSC observations. We also removed sources with $m_i>$25.5 in the HSC catalog because faint sources with $m_i>$25.5 are likely beyond the detection capability of \NuSTAR-NEP survey at $F_{3-8}$ = 8 $\times$ 10$^{-15}$ erg\,cm$^{-2}$\,s$^{-1}$. The HSC observation covered the entire \NuSTAR\ observed field. SDSS is able to detect bright sources ($m_i<$19) with accurately measured magnitude, while HSC is capable of detecting faint sources.

We only matched the optical catalogs with the 22 \NuSTAR\ sources that have soft X-ray counterparts and thus better positional accuracy. The optical and IR properties of the rest 11 soft X-ray non-detected sources will be studied when further soft X-ray data are available. If the \NuSTAR\ detected sources have \cha\ counterparts, the \cha\ detected coordinates were used to match the optical and IR catalogs using a 3\arcsec\ search radius \citep[see, e.g., Fig.~2 in][]{Marchesi_2016}. Otherwise, the \XMM\ detected coordinates were used to match the optical and IR catalogs with a \ang[angle-symbol-over-decimal]{;;4.5} search radius \citep[corresponding to separations between the \XMM\ detected coordinates and optical and IR associations' coordinates of 95\% of the \XMM\ STRIPE 82 sources;][]{LaMassa_2016}. Twenty \NuSTAR\ sources (except ID 1 and ID 29) have at least one optical counterpart within the search radius of soft X-ray detected coordinates. Fourteen sources have a unique optical counterpart. ID 31 has two soft X-ray counterparts, but only one of them has an optical source within the search radius. ID 16, 21, and 27 have two or three optical sources (only one of them is detected by SDSS) within the search radius. We assume that the SDSS detected sources are the most likely optical counterparts of ID 16, 21, and 27 because they are magnitudes brighter than the other HSC detected sources and are the closest to the soft X-ray position\footnote{The distances between soft X-ray positions and SDSS positions of the three sources are \ang[angle-symbol-over-decimal]{;;0.3}, \ang[angle-symbol-over-decimal]{;;1.1}, and \ang[angle-symbol-over-decimal]{;;2.6}, respectively. The distances between soft X-ray positions and HSC positions of the three sources are \ang[angle-symbol-over-decimal]{;;2.5}, \ang[angle-symbol-over-decimal]{;;2.5}, and \ang[angle-symbol-over-decimal]{;;3.3} \& \ang[angle-symbol-over-decimal]{;;4.5} (ID 27 have two HSC detections within the searching radius), respectively.}.

\begin{figure*} 
\begin{multicols}{2}
\includegraphics[width=0.5\textwidth]{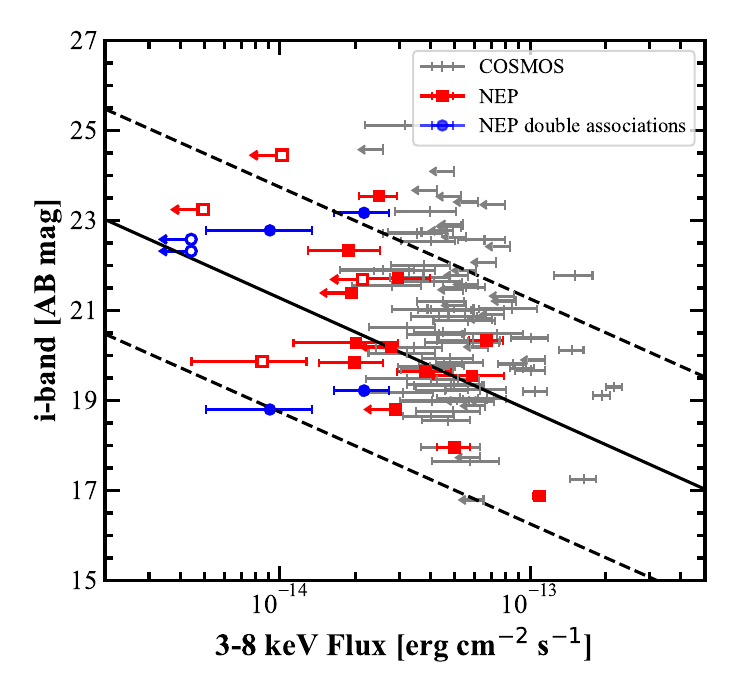}\par
\includegraphics[width=0.5\textwidth]{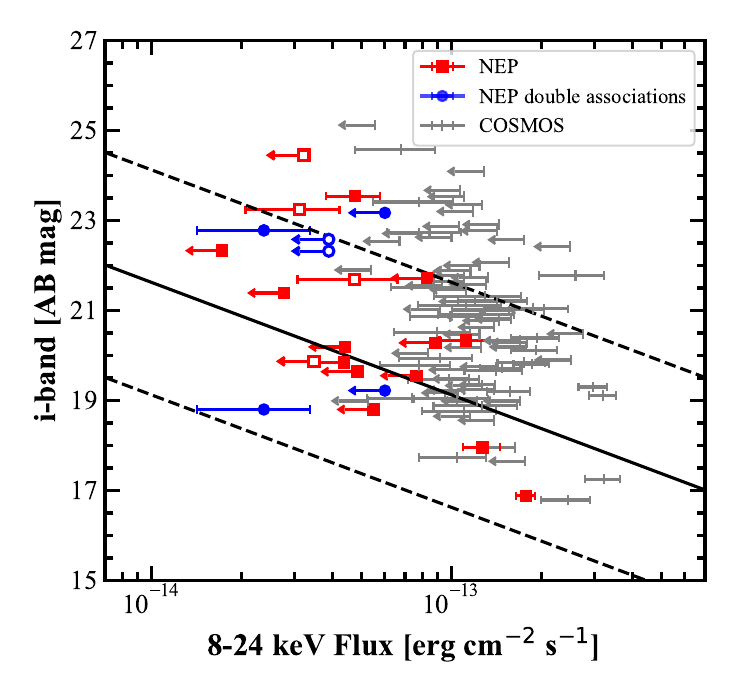}\par
\end{multicols}
\vspace{-0.3cm}
\caption{The $i$-band magnitude as a function of X-ray flux (left: 3--8\,keV; right: 8--24\,keV) of our sample. Sources with single optical counterparts are plotted in red squares and sources with double optical associations are plotted in blue circles. Candidate CT-AGN are labeled as open squares or circles. The solid and dashed lines represent the classical AGN locus \citep{Maccacaro1988} when $X/O$ = 0 $\pm$ 1. We also compare the \NuSTAR\ COSMOS sample \citep[grey;][]{Civano_2015}.}
\label{fig:X_O}
\end{figure*}

ID 6 has one HSC source detected within the search radius with $m_i$ = 23.17 which is $D$ = \ang[angle-symbol-over-decimal]{;;1.6} away from the soft X-ray detected coordinate. However, a bright SDSS source with $m_i$ = 19.22 is also close to the soft X-ray detected coordinate with $D$ = \ang[angle-symbol-over-decimal]{;;3.6}. Although the SDSS source is outside the search radius from the soft X-ray detected coordinate, it is closer to the \NuSTAR\ detected coordinate. Therefore, we associated both optical sources with ID 6 and report both in the catalog. ID 9 has two optical sources within the search radius: a SDSS source with $m_i$ = 18.80 and $D$ = \ang[angle-symbol-over-decimal]{;;2.8}, and a HSC source with $m_i$ = 22.78 and $D$ = \ang[angle-symbol-over-decimal]{;;0.6}. We cannot distinguish the optical counterpart of ID 9 because the brighter one is further from the soft X-ray detected coordinate. Therefore, we consider both optical sources as possible counterparts of ID 9. ID 33 has two HSC sources within the search radius with $m_i$ = 22.32 and 22.58 and $D$ = \ang[angle-symbol-over-decimal]{;;2.2} and \ang[angle-symbol-over-decimal]{;;2.6}, respectively. Therefore, we associate both HSC sources with ID 33. 

No optical sources are within the search radius of the soft X-ray counterparts of ID 1 and 29. However, more than 10 SDSS and HSC sources are within their 20\arcsec\ of the \NuSTAR\ detected coordinate. Therefore, we did not associate any optical source with ID 1 and 29 for now. The detection of ID 1 and 29 in the 2--10 keV band by \XMM\ is $\sim$3\,$\sigma$. A better measurement of the X-ray coordinates of ID 1 and 29 will be obtained after all accepted \NuSTAR\ (for Cycle 6), \XMM\ (together with Cycle 6 \NuSTAR\ observations), and \cha\ (Maksym in prep.) observations are completed and analyzed. Thus may match more optical counterparts using their better measured soft X-ray coordinates. Another possibility is that the optical counterpart of ID 29 is fainter than the optical catalog limit based on its faint soft X-ray flux and the X-ray to optical relation discussed in Section~\ref{sec:X_O}.
 
Therefore, seventeen sources have a single optical counterpart, and three sources have double optical counterparts. Sixteen sources have SDSS counterparts, and six (ID 4, 6, 7, 17, 22, and 24) of them have SDSS measured photometric redshifts. ID 18 has SDSS spectroscopically measured redshift. ID 16 has DBSP spectroscopically measured redshift. 

\subsection{Infrared counterparts}
We searched for IR counterparts to the soft X-rays detected \NuSTAR\ sources using the catalog produced by C. Willmer using {\it MMT} Magellan infrared spectrograph \citep[MMIRS;][]{McLeod2012}. The MMIRS observations covered a region of 15\arcmin\ $\times$ 15\arcmin\ in the {\it JWST} NEP Time-Domain Field, which covered $\sim$40\% of the \NuSTAR\ observations. The observations were taken using four MMIRS filters ($YJHK$). 14 out of the 20 sources that have optical counterparts have IR counterparts, and the coordinates of the IR counterparts are consistent with those of the optical counterparts. The six sources which do not have IR counterparts are out of the MMIRS FoV. 

Five sources (ID 4, 12, 13, 14, and 22) have {\it MMT}/Binospec \citep[0.39--1\,$\mu$m;][]{Fabricant_2019} spectroscopically measured redshifts. Therefore, 11 sources in total have spectroscopic and/or photometric redshifts. The SDSS photometrically measured redshifts of ID 4 and 22 are consistent with their {\it MMT}/Binospec spectroscopically measured redshifts. Based on the available optical and IR spectra, six out of seven sources are determined as AGN with one source being ambiguous. A detailed analysis of the optical and IR properties of the 33 sources will be studied when all of their optical/IR spectra are available in the future work. The MMIRS coordinates and {\it MMT}/Binospec measured redshifts are reported in Table~\ref{Table:catalog}.

We also matched with the AllWISE catalog using the Wide-field Infrared Survey Explorer mission \citep[$WISE$;][]{wright10}. The nominal PSF of $WISE$ is $\approx$ 8\arcsec, which is larger than the matching radii of the \cha\ and \XMM. Therefore, it was used to match the soft X-rays detected coordinates for both \cha\ and \XMM. 18 out of the 20 sources that have optical counterparts have $WISE$ counterpart with only 3 sources having separations $D>$3\arcsec. ID 11 and 31, which lack $WISE$ counterparts, have both optical ($m_i$ = 25.3 and 21.7) and MMIRS (with K-band magnitudes of $m_K$ = 21.6 and 21.3) counterparts, suggesting that they are beyond the sensitivity of $WISE$. The $WISE$ coordinates of the 18 sources are reported in Table~\ref{Table:catalog}.

\subsection{X-ray to optical properties}\label{sec:X_O}
The X-ray--optical flux ($X/O$) ratio has been historically used to study the nature of X-ray detected sources \citep[e.g.,][]{Maccacaro1988}. The $X/O$ ratio is defined as $X/O$ = log($f_{\rm X}/f_{\rm opt}$) = log($f_{\rm X}$) $+$ $m_{\rm opt}$/2.5 $+$ $C$, where $f_{\rm X}$ is the X-ray flux in a given energy band, $m_{\rm opt}$ is the optical magnitude at a chosen band, and $C$ is a constant depending on the energy bands chosen in the X-ray and optical observations. The $X/O$ relation was initially studied using the soft X-ray fluxes, finding that most of the detected AGN in {\it{Einstein}}, {\it{ROSAT}}, and {\it{ASCA}} surveys can be characterized by $X/O$ = 0 $\pm$ 1 \citep{Stocke1991, Schmidt1998, Akiyama_2000}. 

We plotted the $i$-band (SDSS if the optical counterpart has SDSS measurement, otherwise {\it Subaru}/HSC measurement) magnitudes as a function of soft (3--8\,keV) and hard (8--24\,keV) X-ray fluxes in Fig.~\ref{fig:X_O}. The $X/O$ ratios were calculated using constants of $C_{\rm i,3-8}$ = 5.49 and $C_{\rm i,8-24}$ = 5.35 for soft and hard band, respectively. The constants were computed taking into account the optical band used in NEP.  
Sources (ID 1 and 29) that do not have associated optical counterparts were not plotted in Fig.~\ref{fig:X_O}. ID 6, 9, and 33 have double optical associations, so we plotted the $i$-band magnitudes of both optical associations. The $i$-band magnitude of ID 8 measured by SDSS has uncommonly large uncertainties with $m_i$ = 23.38 $\pm$ 0.97, therefore, we use the HSC measured $i$-band magnitude for ID 8 which is $m_i$ = 22.33.

Most of our sources lie in the --1$<X/O<$1 region, suggesting that they are AGN, which is consistent with what is found using their optical/IR spectra. High $X/O$ ratio is found in ID 12 ($z$ = 0.8846), which has $f_{\rm X}/f_{\rm opt}$ = 15.1$_{-4.7}^{+5.1}$ and 27.8$_{-5.5}^{+5.8}$ in 3--8\,keV and 8--24\,keV band, respectively. Another high $X/O$ source (ID 11), which is also the optically faintest one in the sample, is a CT-AGN candidate. High ($>$10) $X/O$ ratio sources were detected in previous \cha\ and \XMM\ surveys \citep{Hornschemeier_2001, Fiore03, Civano05, Brusa_2007, Laird_2008, Xue_2011} and \NuSTAR\ surveys \citep{Civano_2015, Lansbury_2017}, which were associated with high redshift or large obscuration. We note that four out of five CT-AGN candidates (see Section~\ref{sec:HR}) which have optical counterparts have $X/O>$0. We also compared our results with the \NuSTAR\ COSMOS sample \citep{Civano_2015} in Fig.~\ref{fig:X_O}. Both surveys showed that the majority of the \NuSTAR\ detected sources are AGN with --1$<X/O<$1. The two surveys probe different parameter spaces on the X-ray--optical plane with NEP probing the lower X-ray flux regime while COSMOS probed the higher X-ray flux regime.

\begin{figure} 
\centering
\includegraphics[width=.5\textwidth]{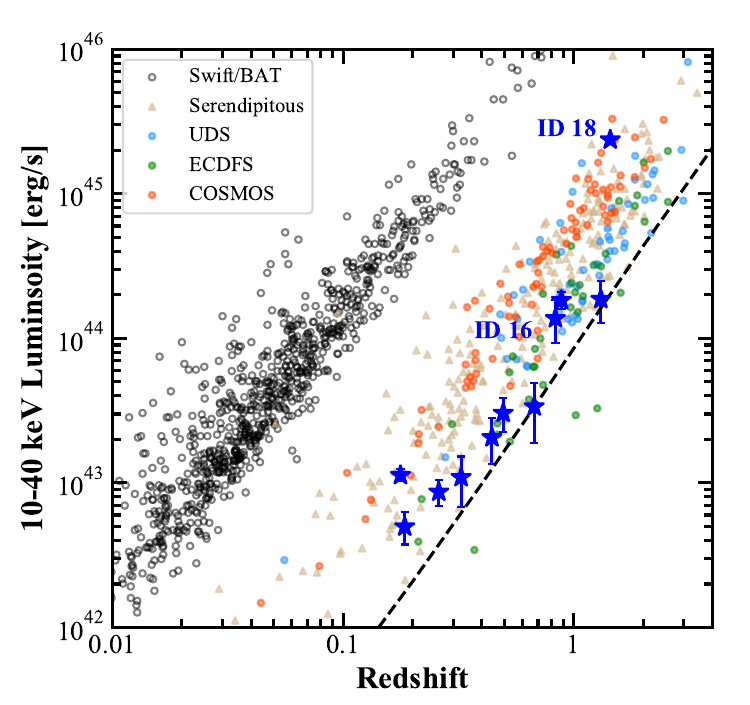}
\caption{10--40\,keV rest frame luminosity versus redshift of the NEP \NuSTAR\ sources (blue star). ID 18 (NLSy1) and ID 16 (\cha\ detected transient) are labeled. The NEP survey sensitivity at 20\% of the sky coverage is plotted as dashed line. \NuSTAR\ COSMOS \citep[red circles;][]{Civano_2015}, ECDFS \citep[green circles;][]{Mullaney15}, UDS \citep[blue circles;][]{Masini_2018}, and serendipitous \citep[brown triangles;][]{Lansbury_2017} surveys are presented as well. We also compare with the {\it{Swift}}-BAT 105-month sample \citep[black open circles;][]{Oh_2018}. The luminosities were not corrected for {any intrinsic} absorption.}
\label{fig:Lx_z}
\end{figure}   

\subsection{Luminosity--Redshift Distribution} \label{sec:Lx-z}
In Fig.~\ref{fig:Lx_z}, we plotted the 10--40\,keV rest-frame luminosities of the 11 \NuSTAR\ NEP sources with redshift information as a function of their redshifts. The 10--40\,keV rest-frame luminosities were computed by converting the 3--24\,keV fluxes assuming a power law model with $\Gamma$ = 1.80 and taking into account the different bandpasses due to the redshift effect. The plotted luminosities were not corrected for any intrinsic absorption. The 3--24\,keV survey sensitivity at 20\% of the sky coverage is plotted. We detected a high-luminosity (10--40\,keV flux of 2.4$\pm$0.1 $\times$ 10$^{45}$\,erg\,s$^{-1}$) and high-redshift source (ID 18, a narrow-line Seyfert 1 galaxy, NLSy1, $z$ = 1.44) in the \NuSTAR\ NEP survey. In Fig.~\ref{fig:Lx_z}, we also present sources detected in previous \NuSTAR\ surveys including the COSMOS \citep{Civano_2015}, ECDFS \citep{Mullaney15}, UDS \citep{Masini_2018}, and 40-month serendipitous \citep{Lansbury_2017} surveys. 99\% of the detected sources from previous \NuSTAR\ surveys were sampled above the NEP sensitivity line, reflecting that the NEP survey indeed reached the deepest sensitivity. 

We also compared \NuSTAR\ detected sources with the sources in the 105-month {\it{Swift}}-BAT catalog \citep{Oh_2018} in Fig.~\ref{fig:Lx_z}. The 10--40\,keV luminosities of the BAT detected sources were converted from the 14--195\,keV luminosities reported in the {\it{Swift}}-BAT catalog assuming a power law model with $\Gamma$ = 1.80. The {\it{Swift}}-BAT and \NuSTAR\ surveys are highly complementary. {\it{Swift}}-BAT samples large sky-area but with low sensitivity, while \NuSTAR\ samples small areas but with high sensitivity. {\it{Swift}}-BAT sampled AGN mostly in the local Universe (with median redshift at $\left<z_{\rm BAT} \right>$ = 0.044), while \NuSTAR\ sampled higher redshift ($\left<z_{\rm NuS} \right>$ = 0.734) sources.

\subsection{Hardness Ratio} \label{sec:HR}
\begin{figure} 
\centering
\vspace{0.2cm}
\includegraphics[width=.5\textwidth]{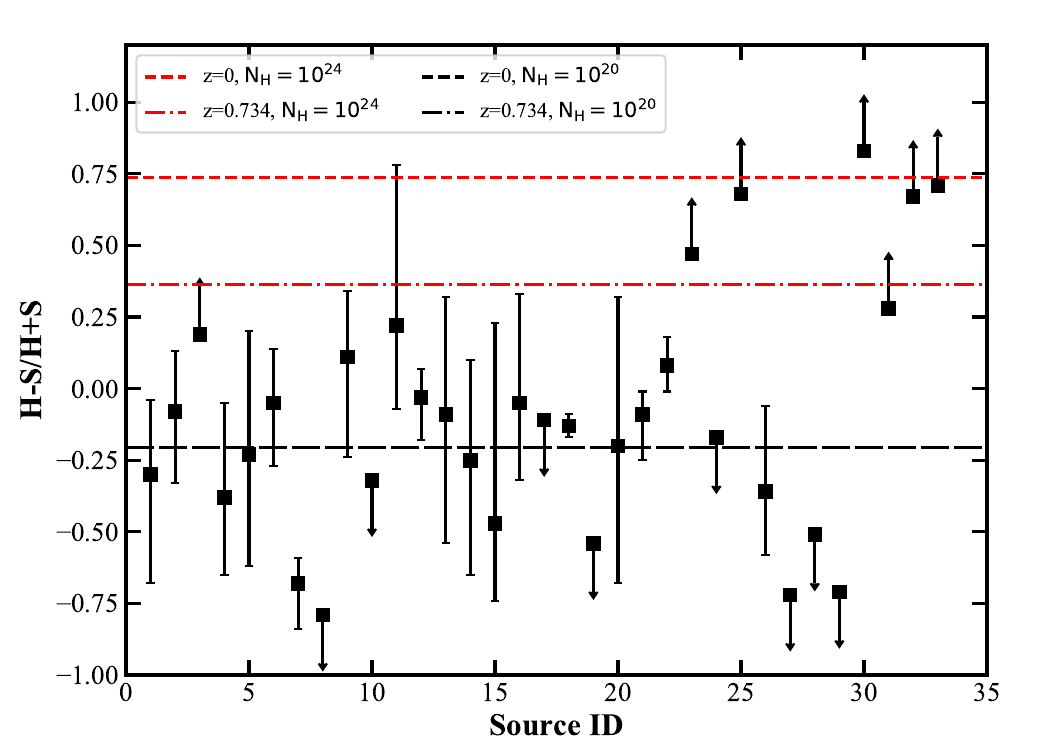}
\includegraphics[width=.5\textwidth]{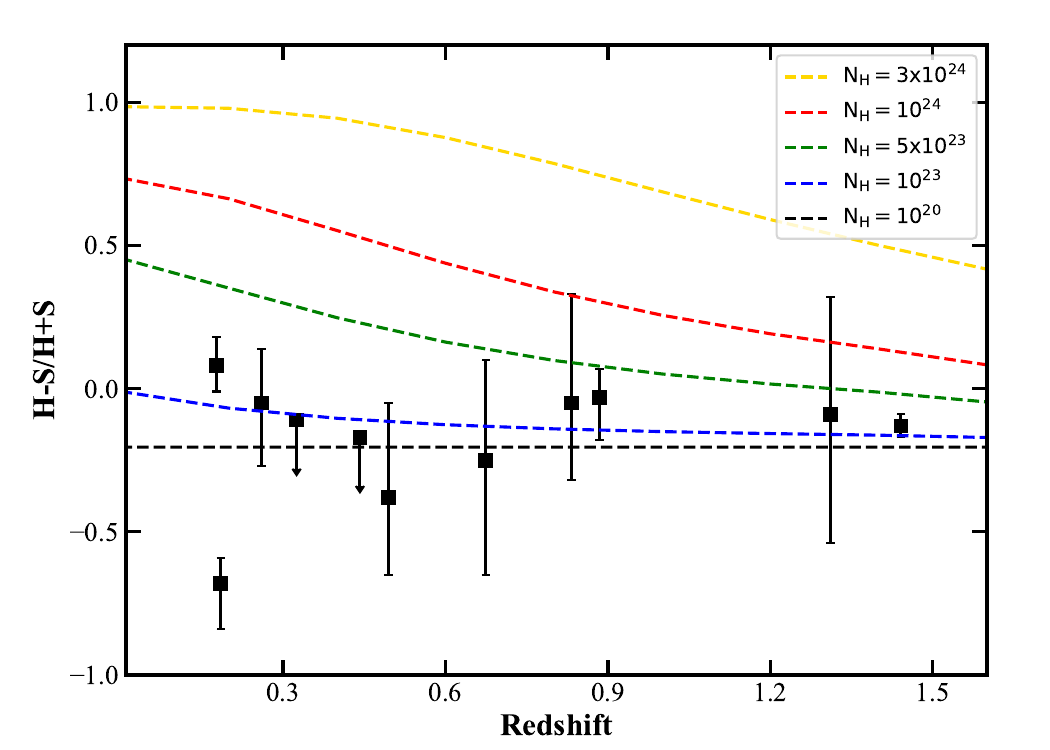}
\caption{Upper: Hardness ratios (with 1\,$\sigma$ errors) of the 33 sources in the 95\% reliability catalog. Expected values of spectral models of N$\rm _{H}$ = 10$^{24}$\,cm$^{-2}$ (red) and 10$^{20}$\,cm$^{-2}$ (black) at $z$ = 0 (dashed line) and $z$ = 0.734 (dash-dot line) are plotted. The expected values of 10$^{20}$\,cm$^{-2}$ at two redshifts are similar, so that the two lines overlap. Lower: Hardness ratio (with 1\,$\sigma$ errors) as a function of redshift. Dashed lines show the expectations using absorbed power law models with $\Gamma$ = 1.80 and different intrinsic column densities of 3 $\times$ 10$^{24}$, 10$^{24}$, 5 $\times$ 10$^{23}$, 10$^{23}$, and 10$^{20}$ cm$^{-2}$ (from top to bottom) as a function of redshift.}
\label{fig:HR}
\end{figure}   

While detailed spectral analysis will be reported in a further publication, to give an estimate of the obscuration of the \NuSTAR\ NEP sample, we computed the hardness ratio of each detected source in the sample using the Bayesian estimation of hardness ratios method \citep[BEHR;][]{Park_2006}. Here, the hardness ratio is defined as HR = (H-S)/(H+S), where H and S are the net counts in the 8--24\,keV and 3--8\,keV bands, respectively. The BEHR method calculates the mean, median, and mode location of the hardness ratio distribution for each source and the uncertainties of these values (we report the mode value and its uncertainty in the catalog), especially for astrophysical sources in the Poisson regime of low counts. The 1\,$\sigma$ error of the hardness ratio is calculated either by Gaussian quadrature numerical integration method if the number of net counts of the hard or soft energy bands is less than 15, or by the Gibbs sampler method for a larger number of net counts. The differences in the effective exposure time between the two energy bands were also considered in the calculation. 

In the upper panel of Fig.~\ref{fig:HR}, the hardness ratio of 33 sources are compared to the expected values of spectral models. The model assumes an absorbed power-law with slope of $\Gamma$ = 1.80 and intrinsic absorption of N$\rm _{H}$ = 10$^{20}$ and 10$^{24}$\,cm$^{-2}$ separately. The models are computed assuming different redshifts: $z$ = 0, and $z$ = 0.734, which is the median redshift of the sources detected in previous \NuSTAR\ surveys, see Section~\ref{sec:Lx-z}. We note that the upper limits of the hardness ratio of seven sources are below the N$\rm _{H}$ = 10$^{20}$\,cm$^{-2}$ model predictions, which might suggest significant soft excess in these sources. The hardness ratio is sensitive to obscuration of N$\rm _{H}$ $>$10$^{23}$\,cm$^{-2}$ \citep[e.g.,][]{Masini_2018}. The relation between the hardness ratio and the obscuration of the source depends on the redshift. Therefore, we plotted the hardness ratio of the 11 sources which have single spectroscopically or photometrically measured redshifts as a function of their redshifts in the lower panel of Fig.~\ref{fig:HR}. We also plotted ID 6 in the lower panel of Fig.~\ref{fig:HR} assuming the redshift measured in one of its double optical and IR counterparts. The hardness ratio values of each source is reported in Table~\ref{Table:catalog}.

ID 30 has a hardness ratio lower limit of 0.83, which is above the CT threshold even for $z$ = 0. Therefore, ID 30 is a candidate CT-AGN despite lacking redshift information. For five sources without redshift (ID 3, 23, 25, 31, 32, and 33), their hardness ratio are above the CT threshold assuming a median redshift $z$ = 0.734, making them CT-AGN candidates. The hardness ratio upper limits of ID 11 (even for $z$ = 0) and ID 13 ($z$ = 1.311) are above the CT threshold, as shown in the lower panel of Fig.~\ref{fig:HR}. Therefore, they are also CT-AGN candidates. Details about these CT-AGN candidates and the CT-AGN fraction are discussed in Section~\ref{sec:CT}.

\begin{figure} 
\centering
\vspace{0.2cm}
\includegraphics[width=.5\textwidth]{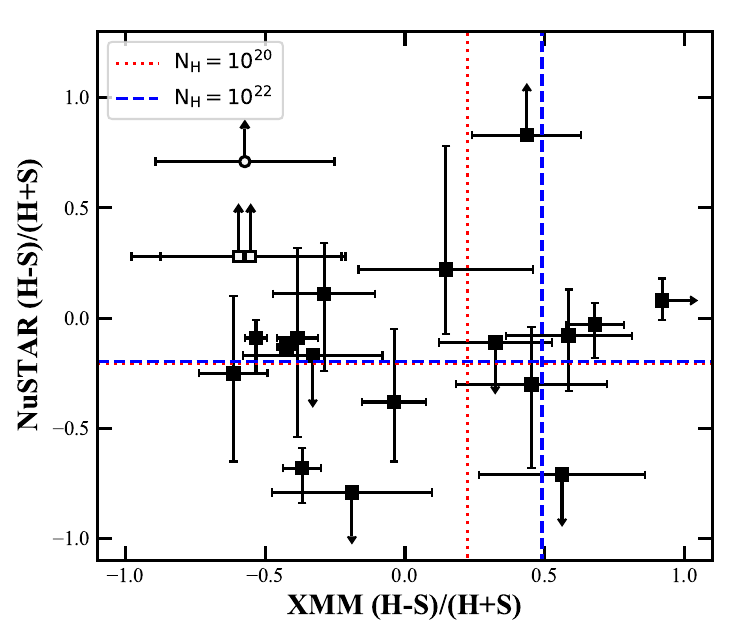}
\caption{\NuSTAR\ hardness ratio (with 1\,$\sigma$ errors) as a function of their \XMM\ counterpart hardness ratio (with 1\,$\sigma$ errors). Expected hardness ratio values of spectral models of N$\rm _{H}$ = 10$^{20}$\,cm$^{-2}$ (red dotted line) and 10$^{22}$\,cm$^{-2}$ (blue dashed line) at $z$ = 0.734 of \NuSTAR\ and \XMM\ are plotted. ID 31 has two \XMM\ counterparts, and they are plotted in open squares. ID 33 is plotted in open circle.}
\label{fig:HR_xmm}
\end{figure}

We computed the hardness ratio of \XMM\ detected sources using the 2--10\,keV and 0.5--2\,keV bands fluxes. The hardness ratios measured in \XMM\ are compared with those measured in \NuSTAR\ in Fig.~\ref{fig:HR_xmm}. ID 31 has two \XMM\ counterparts, so both of them are plotted in the figure. Most of the sources which are measured as obscured or unobscured AGN by \NuSTAR\ are also measured as obscured or unobscured by \XMM\ within uncertainties (in the upper right and lower left corners of Fig.~\ref{fig:HR_xmm}). ID 33 is detected as heavily obscured AGN by \NuSTAR\ but is determined as unobscured AGN by \XMM\ in the upper left of the figure. However, ID 33 could still be considered a possible CT-AGN candidate, given the simple spectral model used in categorizing the sources. For instance, ID 33 might have strong non-AGN thermal emissions at $<$2\,keV, so that its \XMM\ hardness ratio is significantly soft. Follow-up spectral analysis of the sources with enough counts will be used to determine the spectral parameters (slope and obscuration at minimum). 

\section{Discussion}\label{sec:discuss}
\subsection{CT Fraction} \label{sec:CT}
As discussed in Section~\ref{sec:HR}, ID 30 is a good CT-AGN candidate. Therefore, the lower limit of the CT-AGN fraction of the NEP 95\% catalog is 1/33 $\sim$3\%. The mode values of hardness ratio of five sources are above the CT threshold if their redshifts are assumed to be $z$ = 0.734. Thus, the expected CT fraction of the catalog is 7/33 $\sim$21\%. The hardness ratio upper limits of the two sources are above the CT threshold. Therefore, the upper limit of CT-AGN fraction is 9/33 $\sim$27\%. As a result, the observed CT fraction of the NEP 95\% reliability catalog is 21$^{+6}_{-18}$\% at 1\,$\sigma$ confidence level. The CT fraction of the NEP survey will be better constrained when all sources have redshift measurements with the future {\it JWST} and {\it J-PAS} observations.
We note that the nine CT candidates were not significantly detected in the \NuSTAR\ soft (3--8\,keV) band (except for ID 13) and they were neither detected by \XMM\ (except for ID 30). 

The observed CT fraction measured in the NEP field is consistent with the CT fraction measured in the \NuSTAR\ COSMOS sample \citep[13\%--20\%;][]{Civano_2015}, which was estimated using the hardness ratio as well. \citet{Masini_2018} computed the CT fraction of the \NuSTAR\ UDS sample, which is 11.5\%$\pm$2.0\% by performing spectral analysis. The result in the UDS field is in agreement with the lower limit of the NEP and COSMOS field, although they used different methods when estimating the N$\rm _H$ values. \citet{Masini_2018} reported that the CT fraction measured in the conservative sample (99\% reliability catalog, $f_{\rm CT,99}\sim$3\%$\pm$2\%) is significantly lower than their 97\% reliability catalog. However, this significant drop of the CT fraction is not found in the NEP field, where the CT fraction in our 99\% reliability catalog is [5\%--24\%]. \citet{Burlon11} and \citet{Ricci15} found a CT fraction of $\sim$4.6\%--7.6\% using the {\it{Swift}}-BAT sample. A recent work \citep{Nuria}, which analyzed the CT-AGN candidates in the BAT sample using high-quality \NuSTAR\ observations, found a lower CT fraction of the BAT sample of 3.5\%$\pm$0.5\%. 

\begin{figure} 
\centering
\includegraphics[width=.5\textwidth]{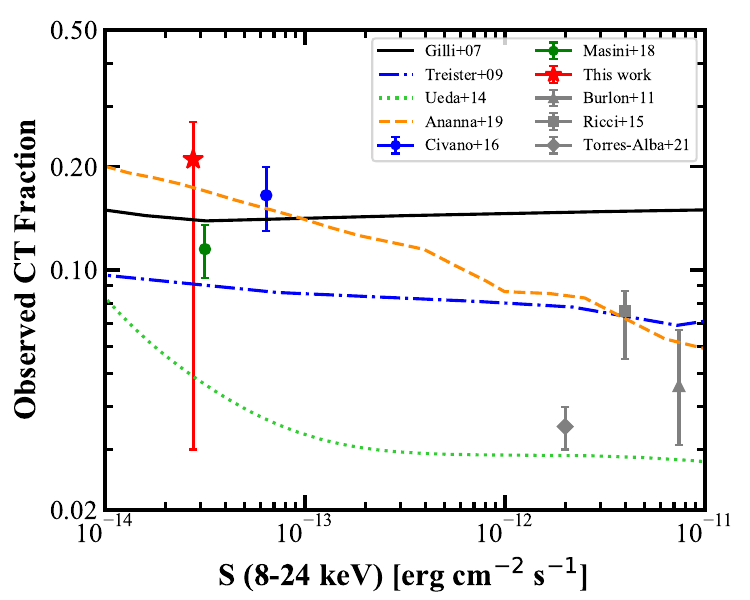}
\caption{Observed CT fraction as a function of 8--24\,keV flux. The CT fraction measured by \NuSTAR\ in NEP field is plotted as a red star at the 8--24\,keV survey sensitivity at 20\% of the sky coverage. Blue and green circles are the \NuSTAR\ measurements in COSMOS \citep{Civano_2015} and UDS \citep{Masini_2018} fields. Gray triangle, square, and diamond are the {\it{Swift}}-BAT measurements by \citet{Burlon11}, \citet{Ricci15}, and \citet{Nuria}. We compared with the CT fractions predicted by the CXB population synthesis models with \citet{gilli07} (black solid line), \citet{Treister09} (blue dash-dot line), \citet{Ueda14} (green dot line), and \citet{Tasnim_Ananna_2019} (orange dashed line).}
\label{fig:CT}
\end{figure}   

The observed CT fraction obtained in this work and the other \NuSTAR\ and {\it{Swift}}-BAT measurements are plotted in Fig.~\ref{fig:CT}. We also compare with the CT fractions predicted by the population synthesis model of CXB \citep{gilli07, Treister09, Ueda14, Tasnim_Ananna_2019} in Fig.~\ref{fig:CT}. The recent \citet{Tasnim_Ananna_2019} model are in good agreement with the hard X-ray observed CT-AGN fraction at both bright and faint fluxes. 

We detected one good CT-AGN candidate (ID 30) and two possible CT-AGN candidates (ID 13 and 31) in the 8--24\,keV band, which are above the 95\% reliability threshold. This is consistent with the CXB population synthesis models, which predicted 2 \citep{gilli07} or 0.7 \citep{Ueda14} CT-AGN being detected in the 8--24\,keV band, respectively. The CT fraction reported in this work is the observed one. An estimation of the intrinsic CT fraction based on our sample is beyond the scope of this paper and further X-ray spectral analysis as well as a large number of photometric and spectroscopic redshifts will allow a better estimate of obscuration and CT fraction. The spectrum of each source will be analyzed using Markov chain Monte Carlo (MCMC) parameter estimation techniques as described in \citet{Lanzuisi2018} in a further work and the full obscuration probability distribution function (PDF) will be used to derive the true CT fraction.

\subsection{Variability analysis} \label{sec:var}
Variability of high redshift ($z>$0.5) AGN has been studied in deep \cha\ and \XMM\ extragalactic surveys \citep[e.g.,][]{Lanzuisi_2014, Yang_2016, Paolillo_17}. However, such studies have not been performed in hard X-ray band. The $JWST$ NEP field is a new effort to study the time-domain astrophysics and the \NuSTAR\ NEP survey was designed to have variability as its primary focus. The field was observed in three epochs during October 2019, January 2020, and March 2020. 

Among the 33 detected sources, 8 sources were observed only in one epoch, 14 sources were observed in two epochs, and 11 sources were observed in all three epochs. To study the variability properties of the 25 sources observed in multiple epochs, we computed their vignetting corrected count rates as a function of time in five energy bands. The observation data (FPMA+FPMB), simulated background, and exposure map of the three observations in a given epoch were merged into mosaics. We extracted the total and background counts on the observation and background mosaics at the coordinate of each \NuSTAR\ detection using a circular aperture of 20\arcsec\ radius in each epoch. The count rates were then computed by dividing the net counts by the vignetting-corrected exposure time in a given energy band for each epoch. Therefore, the count rate has taken into account the variations in the vignetting correction among different epochs of observations. The net count errors were computed using Equations (9) and (12) in \citet{Gehrels1986} at 90\% confidence level. 

The variability is determined by using the differences of the count rates of the source in different epochs and compare these with the combined errors of the count rates in different epochs (e.g., Err$_{cob}$ = (Err$_1^2$+Err$_2^2$)$^{1/2}$). At 90\% confidence level, six sources (ID 3, 4, 6, 13, 19, and 20) exhibited only short-term variability between epoch one and epoch two (two months) or/and between epoch two and epoch 3 (three months) in 3--24\,keV. Six sources (ID 15, 16, 22 ,23, 24, and 29) exhibited only long-term variability between epoch one and epoch three (five months). Four sources (ID 12, 17, 18, and 32) exhibited both long-term and short-term variabilities. Therefore, among the 25 sources observed in multiple epochs, 16 sources showed variability in the short-term or/and long-term. This is in agreement with what was found in the 15 years \cha\ Deep Field-South observations that 74\% of the sources in the survey showed flux variability \citep{Yang_2016}. 

\begin{figure} 
\centering
\includegraphics[width=.5\textwidth]{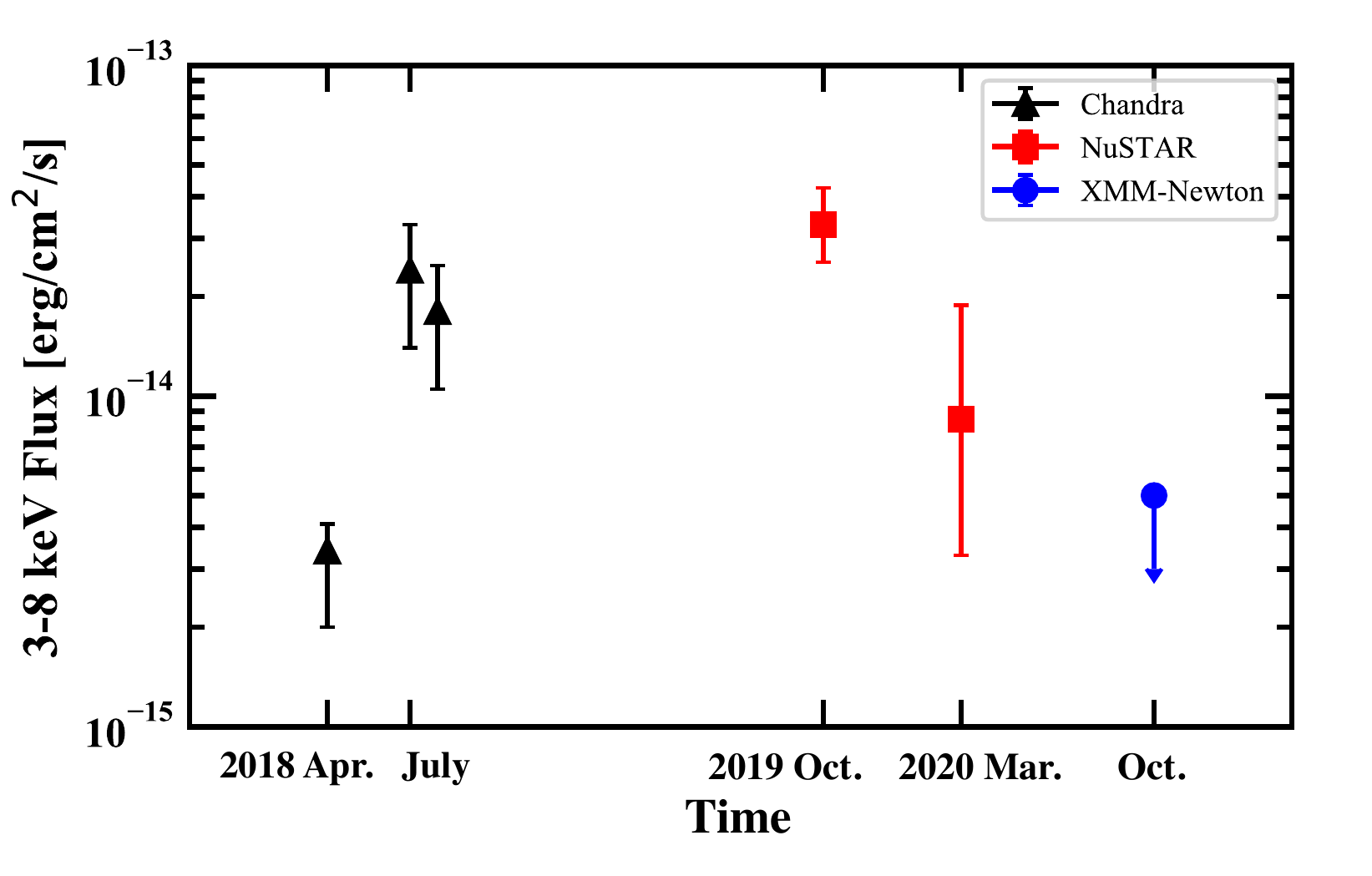}
\includegraphics[width=.5\textwidth]{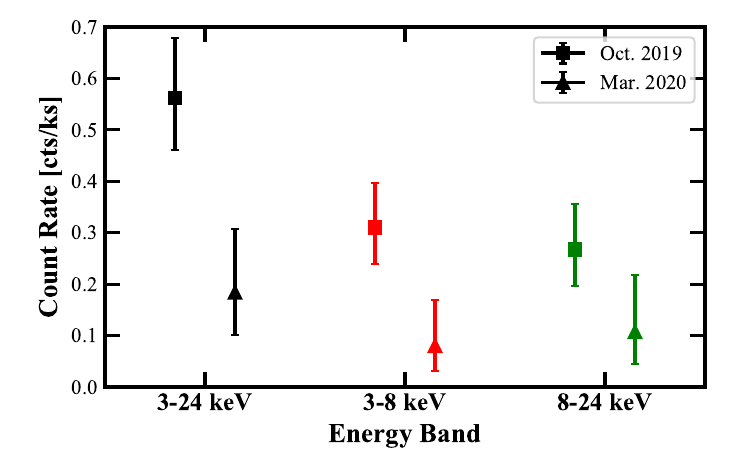}
\caption{Upper: 3--8\,keV flux of ID 16 (SDSS J172421.74+ 654847.5) as a function of time. \cha\ and \NuSTAR\ observations are labeled using black triangles and red squares, respectively. \XMM\ didn't detect the source in Oct. 2020, so it is plotted as the upper limit (blue circle). The flux uncertainties are at 90\% confidence level. Lower: Vignetting corrected \NuSTAR\ count rates of ID 16 in the two epochs (square for epoch 1 and triangle for epoch 2) in three energy bands.}
\label{fig:ID16}
\end{figure}

SDSS J172421.74+654847.5 (ID 16) is a variable AGN found in \cha\ observations, whose flux was found to increase by a factor of $\sim$7 from April 11 to July 26 in 2018 but then decreased by 25\% on July 31 (ATel \# 11906 and ATel \# 12049). The source flux was computed by fitting its \cha\ spectra. \NuSTAR\ observed this source twice in 2019 October and 2020 March. We found that the 3-8\,keV flux of this source was higher than the \cha\ detected flux in 2019 October, but decreased by $\sim$74\% in 2020 March. The \NuSTAR\ flux was computed by converting its count rates measured in each epoch using the count-rate-to-flux CF (3.39 $\times$ 10$^{-11}$\,erg\,cm$^{-2}$ counts$^{-1}$). We present the 3--8\,keV flux as a function of time in the upper panel of Fig.~\ref{fig:ID16}. \XMM\ did not detect this source, so we plot an upper limit of flux 8 $\times$10$^{-15}$\,erg\,cm$^{-2}$\,s$^{-1}$ (2--10\,keV flux limit of the \XMM\ detection). In the lower panel of Fig.~\ref{fig:ID16}, we also presented its count rates in the two \NuSTAR\ epoch in 3--24\,keV, 3--8\,keV, and 8--24\,keV bands. We also found an increasing X-ray absorption in ID 16 at $>1\,\sigma$ confidence level. The HR of ID 16 in \NuSTAR\ epoch 1 is $-0.05^{+0.32}_{-0.37}$ and $0.42^{+0.20}_{-0.10}$ in epoch 3. This significant variability of its HR suggests that X-ray absorption might play a significant role in the flux drop of ID 16. The source will be covered again by \NuSTAR\ in cycle 6, which will enable us to better understand the flux and absorption variability of ID 16.

\subsubsection{Source detection by epoch}
When generating the source catalog, we performed source detection on the mosaics of all three epochs' observations to achieve the deepest exposure. Variable sources, which were luminous in a single epoch, might be missed when performing source detection on the full mosaic. Therefore, to explore the existence of such events, we performed source detection on each epoch in five energy bands on FPMA+FPMB mosaics. Following the same source detection technique in Section~\ref{sec:catalog}, we created the 95\% reliability source catalog for each epoch. We used the same DET\underline{\;\;}ML--reliability relations in different exposure intervals obtained in Section~\ref{sec:relia}. We computed the exposure map for each epoch. The area covered by each epoch are 0.080\,deg$^2$, 0.097\,deg$^2$ and 0.096\,deg$^2$. The deepest vignetting-corrected exposure for each epoch reaches 337\,ks, 351\,ks, and 264\,ks.

In epoch one, fourteen sources were detected above the 95\% reliability threshold. In the area covered by epoch one, twelve sources were detected in the total mosaic but were not detected in epoch one, suggesting that the signal-to-noise (S/N) ratios of these sources were below the detection threshold in a single epoch. One of the sources detected in epoch one (ID E1--12) was not detected in the total mosaic. The effective exposure time of this source is 22\,ks and it was only detected above 95\% reliability threshold in 3--24\,keV (DET\underline{\;\;}ML = 13.10). We did not find any \XMM\ and \cha\ counterparts of ID E1--12. The position of ID E1--12 was also observed in epoch three, but no source was detected within 30\arcsec\ of its position. Therefore, ID E1--12 might be a spurious detection due to background fluctuation, although we cannot exclude the possibility that the source was only bright in epoch one so that we did not detect it in epoch three and \XMM\ and \cha\ observations.

In epoch two, we detected fourteen sources above the 95\% reliability threshold. In the area covered by epoch two, eleven sources were detected in the total mosaic but not in epoch two. One of the sources detected in epoch two (ID E2--14) was not detected in the total mosaic. The effective exposure time of this source is 139\,ks and it was only detected above 95\% reliability threshold in 3--8\,keV (DET\underline{\;\;}ML = 11.87). We did not find any \XMM\ and \cha\ counterparts of ID E1--12. The position of ID E2--14 was also observed in epoch one and three, but no source was detected within 30\arcsec\ of its position. The source was close to (37\arcsec\ separation) the brightest source (ID 18, the NLSy1) in the FoV of the survey. Therefore, ID E2--14 might be a spurious detection since it was on the edge of the PSF of a bright source.

We only detected ten sources above the 95\% reliability threshold in epoch three due to its lower exposure time compared to epochs one and two. In the area covered by epoch three, eleven sources were detected in the total mosaic but not in epoch three. One of the sources detected in epoch three (ID E3--6) was not detected in the total mosaic. The effective exposure time of this source is 165\,ks and it was only detected above 95\% reliability threshold in 3--24\,keV (DET\underline{\;\;}ML = 12.79). We found a soft X-ray counterpart of ID E3--6 detected by \cha\ ($D$ = \ang[angle-symbol-over-decimal]{;;17.3}). \XMM\ also detected a source close to ID E3--6 but with $D$ = \ang[angle-symbol-over-decimal]{;;21.6}. One HSC ($m_i$ = 24.7) source was associated with ID E3--6 using the \cha\ coordinate. The position of ID E3--6 was also observed in epoch one, but it was not detected. Therefore, ID E3--6 might be a variable source, which was dim in epoch one but became luminous in \NuSTAR\ epoch three, \XMM\ and \cha\ observations. 

This preliminary study of source variability among epochs shows that $JWST$ NEP Time-Domain Field is a field rich in variable events. An overall analysis of the variable sources is out of the scope of this paper. We will perform a comprehensive study of the X-ray variability in the $JWST$ NEP Time-Domain Field when all observations are completed including 4 epochs of \NuSTAR\ observations in Cycle 6, together with long term soft X-ray observations in the same field. The study will include but is not limited to continuous monitoring of the discovered variable sources, studying the 3--8\,keV and 8--24\,keV light curves of the sources, and analyzing the hardness ratio variability of each source among different epochs of observations.

\section{Conclusion}
In this paper, we presented the \NuSTAR\ survey of the {\it JWST} NEP Time-Domain Field, which is one of the most sensitive \NuSTAR\ surveys. The initial survey consists of nine observations conducted in three epochs from September 2019 to March 2020 with a total exposure time of 681\,ks and covers an area of $\sim$0.16\,deg$^2$. 
\begin{enumerate}
\item We detected 21 sources above the 99\% reliability threshold and 33 sources above the 95\% reliability threshold in the \NuSTAR\ survey of the {\it JWST} NEP Time-Domain Field. The source detection was performed in six energy bands including 3--24\,keV, 3--8\,keV, 8--24\,keV, 8--16\,keV, 16--24\,keV, and 35--55\,keV. The distribution of the detected sources in the 95\% reliability catalog is listed in Table~\ref{Table:FSH}. Only one source was detected in 16--24\,keV band, and no source was detected in 35--55\,keV band. We focused our analysis on the 95\% reliability catalog, in which we statistically expect only $\sim$2 spurious detections.

\item We computed the log$N$-log$S$ of 3--8\,keV, 8--24\,keV, and 8--16\,keV bands, which are presented in Fig.~\ref{fig:logN-logS}. The lower limits of the 3--8\,keV number counts obtained in our work are consistent with the measurements of previous \NuSTAR\ and soft X-ray surveys and the constraints from the CXB population synthesis models. The number counts of 8--24\,keV and 8--16\,keV bands are in good agreement with previous \NuSTAR\ measurements and the constraints of the CXB population synthesis models.

\item We analyzed two soft X-ray (10\,ks \XMM\ and 20\,ks \cha) observations taken in 2020, which covered about 90\% of the \NuSTAR\ field. We found that 22 \NuSTAR\ sources have a soft X-ray counterpart detected by \XMM\ and/or \cha. The \NuSTAR\ sources which do not have soft X-ray counterpart might be the sources not covered by the soft X-ray observations, faint sources below the detection capability of the soft X-ray observations, variable sources, heavily obscured AGN, or spurious detections.

\item We explored the X-ray-optical properties of the \NuSTAR\ detected sources. The $i$-band magnitude of the sources as a function of their 3--8\,keV and 8--24\,keV fluxes are plotted in Fig.~\ref{fig:X_O}. Most of the detected sources are in the locus of $X/O$ = 0 $\pm$1, suggesting that they are AGN, which is consistent with what is found using their optical and IR spectra. Pervious \NuSTAR\ extragalactic surveys also showed that the detected sources are dominated by AGN, but the NEP survey reaches a deeper X-ray flux.

\item Eleven sources have measured redshifts. Their luminosities versus redshifts are plotted in Fig.~\ref{fig:Lx_z}, showing that the \NuSTAR\ NEP survey is one of the most sensitive survey among \NuSTAR\ contiguous surveys to date. The brightest source detected in the \NuSTAR\ NEP survey is a NLSy1 ($z$ = 1.44) and the spectral analysis of this source alone will be presented in Zhao et al. in prep.

\item We computed the hardness ratio of each source and plotted them in Fig.~\ref{fig:HR}. We found that one sources is a good CT-AGN candidate, and eight sources are possible CT-AGN candidates. The CT fraction of the whole sample is 21$^{+6}_{-18}$\% at 1\,$\sigma$ confidence level, which is consistent with the results of previous \NuSTAR\ surveys. The uncertainties on the CT fraction will be significantly reduced when the redshifts of all sources in the catalog are measured. A broadband X-ray spectral analysis of all detected sources will be presented in a future work.

\item The survey was designed to have variability as its prime focus, and our study has shown that the $JWST$ NEP Time-Domain Field is indeed rich in variable sources. 16 out of the 25 sources observed in multi-cycles showed variabilities at 90\% confidence level in 3--24\,keV band. We also performed source detection by epoch and found three new sources which were not detected in the mosaics combining all observations. A further comprehensive variability study will be conducted after all observations approved in \NuSTAR\ cycle 6 are taken.

\end{enumerate}

\section*{Acknowledgements}
We thank the anonymous referee for their helpful comments. XZ acknowledges NASA funding under contract number 80NSSC20K0043. WPM acknowledges support from Chandra grants GO8-19119X, GO9-20123X and GO0-211126X. DMA, DJR, and MJW thank the Science and Technology Facilities Council for funding through grant codes ST/P000541/1 and ST/T000244/1. RAW acknowledges support from NASA JWST Interdisciplinary Scientist grants NAG5-12460, NNX14AN10G and 80NSSC18K0200 from GSFC. We thank Karl Foster and the \NuSTAR\ observation planning team for their help in designing the observing plan and in scheduling the observations. We thank Daniel Wik and Brian Grefenstette for their discussions and help on {\tt nuskybgd}. We thank Satoshi Kikuta for reducing the {\it Subaru} images. We thank the \NuSTAR\ operations, software and calibrations teams for support with these observations. This research has made use of data and software provided by the High Energy Astrophysics Science Archive Research Center (HEASARC), which is a service of the Astrophysics Science Division at NASA/GSFC and the High Energy Astrophysics Division of the Smithsonian Astrophysical Observatory. This work is based on observations obtained with \XMM, an ESA science mission with instruments and contributions directly funded by ESA Member States and NASA. This research has made use of data obtained from the \cha\ Data Archive, and software provided by the \cha\ X-ray Center (CXC) in the application package CIAO. This work is partly based on observations at the MMT, a joint facility operated by the Smithsonian Institution and the University of Arizona. AllWISE makes use of data from WISE, which is a joint project of the University of California, Los Angeles, and the Jet Propulsion Laboratory/California Institute of Technology, and NEOWISE, which is a project of the Jet Propulsion Laboratory/California Institute of Technology. We acknowledge use of the SMOKA Science Archive, developed and maintained by the Astronomical Data Archives Center (ADAC), Astronomy Data Center (ADC), National Astronomical Observatory of Japan (NAOJ). 

\section*{Data Availability}
The electronic version of the generated 95\% reliability catalog can be found on VizieR (\url{https://cdsarc.cds.unistra.fr/viz-bin/Cat?J/MNRAS/508/5176}). 

\bibliographystyle{mnras}
\bibliography{referencezxr} 

\appendix

\section{Catalog Description}
The description of each column of the catalog is described in Table~\ref{Table:catalog}.
\begingroup
\renewcommand*{\arraystretch}{1.3}
\begin{table*}
\centering
\caption{95\% reliability level catalog descriptions}
\label{Table:catalog}
  \begin{tabular}{ll}
       \hline
       \hline     
 	Col.&Description\\
	\hline 
	1&Source ID.\\
	2&\NuSTAR\ source name, following the standard IAU convention with the prefix ``\NuSTAR''.\\
	3--4&The X-ray coordinates of the source (the position of the source in which energy band has the highest DET\underline{\;\;}ML).\\
	5&The 3--24\,keV band deblended DET\underline{\;\;}ML (--99 if undetected). \\
	6&The 3--24\,keV band vignetting-corrected exposure time in ks at the position of the source.\\
	7&The 3--24\,keV band total counts in a 20\arcsec\ radius aperture.\\
	8&The 3--24\,keV band deblended background counts in a 20\arcsec\ radius aperture (--99 if undetected).\\
	9&The 3--24\,keV band not deblended background counts in a 20\arcsec\ radius aperture.\\
	10&The 3--24\,keV band net counts (deblended if detected \& above DET\underline{\;\;}ML threshold or 90\% confidence upper limit if \\ 
	&undetected or detected but below DET\underline{\;\;}ML threshold) in a 20\arcsec\ radius aperture.\\
	11&The 3--24\,keV band 1\,$\sigma$ positive net counts error (--99 for upper limits).\\
	12&The 3--24\,keV band 1\,$\sigma$ negative net counts error (--99 for upper limits).\\
	13&The 3--24\,keV band count rate (90\% confidence upper limit if not detected or detected but below the threshold) in a 20\arcsec \\
	&radius aperture.\\
	14&The 3--24\,keV band aperture corrected flux (90\% confidence upper limit if not detected or detected but below threshold).\\
	15&The 3--24\,keV band positive flux error (--99 for upper limits).\\
	16&The 3--24\,keV band negative flux error (--99 for upper limits).\\
	17--28&Source properties in 3--8\,keV band with the same order as column 5--16.\\
	29--40&Source properties in 8--24\,keV band with the same order as column 5--16.\\
	41--52&Source properties in 8--16\,keV band with the same order as column 5--16.\\
	53--64&Source properties in 16--24\,keV band with the same order as column 5--16.\\
	65&Hardness ratio computed using BEHR.\\
	66&Upper limit of hardness ratio.\\
	67&Lower limit of hardness ratio.\\
	68,69&Soft X-ray (\cha\ if having \cha\ counterpart, otherwise \XMM) coordinates of the associated source\\
	&(--1 if no soft X-ray counterpart).\\
	70&\NuSTAR\ to soft X-ray counterpart position separation in arcsec.\\
	71&3--8\,keV flux from \XMM\ (90\% confidence upper limit if $likemin<$6).\\
	72&3--8\,keV \XMM\ flux 1$\sigma$ error (-99 for upper limit). \\
     	73,74&Optical coordinates of the associated source.\\
	75&Optical $i$ band magnitude.\\
	76&Flag of Optical $i$ band magnitude (0 = SDSS, 1 = HSC)\\
     	77,78&MMIRS coordinates of the associated source.\\
	79,80&WISE coordinates of the associated source.\\
	81&Spectroscopic redshift of the associated source.\\
	82&Photometric redshift of the associated source.\\
	83&Luminosity Distance in Mpc using the spectroscopic (if have) or photometric redshift.\\
	84&10--40\,keV band rest-frame luminosity.\\
	85&10--40\,keV band positive rest-frame luminosity error.\\
	86&10--40\,keV band negative rest-frame luminosity error.\\
	87&Flag for sources with multiple low energy counterparts (0 = false, 1 = true).\\
	88&Flag for sources with multiple optical and IR counterparts (0 = false, 1 = true).\\
	\hline
\end{tabular}
\end{table*}
\endgroup

\bsp	
\label{lastpage}
\end{document}